\documentclass[aps,prd,twocolumn,floatfix,
10pt,
amssymb,amsfont,amsmath,
superscriptaddress,
float,nofootinbib,longbibliography]{revtex4-2}

\usepackage[usenames,dvipsnames]{xcolor}
\usepackage{graphicx,color,dsfont,xcolor,amsmath, amssymb,braket,yfonts,wasysym,stmaryrd,nicefrac}
\usepackage[export]{adjustbox}
\usepackage{gensymb}
\usepackage{hyperref}
\usepackage[caption=false]{subfig} 
\usepackage[normalem]{ulem}
\usepackage[titletoc,toc,title]{appendix}
\usepackage[all,cmtip]{xy}

\usepackage{tikz}
\usetikzlibrary{external,calc,decorations.pathreplacing,decorations.markings,decorations.pathmorphing,arrows.meta,shapes.geometric,angles,quotes,intersections,snakes}

\DeclareMathOperator{\im}{im}

\graphicspath{{fig/}}

\newcommand{\la}{\langle}
\newcommand{\ra}{\rangle}
\newcommand{\q}{\quad}
\newcommand{\sss}{\scriptstyle}

\newcommand{\nn}{\nonumber}
\newcommand{\mc}{\mathcal}
\newcommand{\apeSet}{\{ a_{\mathsf e , \mathsf p} \}_{\mathsf p \supset \mathsf e}}
\newcommand{\aepSet}{\{ a_{\mathsf e , \mathsf p} \}_{\mathsf e \subset \partial \mathsf p}}

\newcommand{\aep}{a_{\mathsf e , \mathsf p}}

\newcommand{\cubu}{{\mathbb T^3_\boxempty}}
\newcommand{\repA}{2-form }
\newcommand{\repB}{1-form }

\include{TikZ}

\hypersetup{%
	bookmarksopen=true,
	unicode=true,           
	pdftoolbar=true,        
	pdfmenubar=true,        
	pdffitwindow=false,     
	pdfstartview={FitH},    
	pdftitle={On the stability of topological order in tensor network states},
	pdfauthor={},
	pdfsubject={},          
	pdfcreator={},          
	pdfproducer={pdfLaTeX}, 
	pdfkeywords={topological order} {tensor networks} {stability},
	pdfnewwindow=true,      
	colorlinks,
	citecolor=BlueViolet,
	filecolor=BlueViolet,
	linkcolor=BlueViolet,
	urlcolor=BlueViolet,
	linktocpage=true
}

\begin{document}

\title{
 On the stability of topological order in tensor network states
 }
\author{Dominic J. Williamson}
\affiliation{Stanford Institute for Theoretical Physics, Stanford University, Stanford, CA 94305, USA}
\author{Clement Delcamp}
\affiliation{Max-Planck-Institut f{\"u}r Quantenoptik,  Hans-Kopfermann-Stra\ss{}e 1, 85748 Garching, Germany}
\affiliation{Munich Center for Quantum Science and Technology (MCQST), Schellingstra\ss{}e 4, 80799 M{\"u}nchen, Germany}
\author{Frank Verstraete}
\affiliation{Department of Physics and Astronomy, Ghent University, Krijgslaan 281, S9, B-9000 Ghent, Belgium}
\author{Norbert Schuch\vspace{0.8em}}
\affiliation{Max-Planck-Institut f{\"u}r Quantenoptik,  Hans-Kopfermann-Stra\ss{}e 1, 85748 Garching, Germany}
\affiliation{Munich Center for Quantum Science and Technology (MCQST), Schellingstra\ss{}e 4, 80799 M{\"u}nchen, Germany}
\affiliation{University of Vienna, Faculty of Physics, Boltzmanngasse 5, 1090 Wien, Austria}
\affiliation{University of Vienna, Faculty of Mathematics,\unpenalty~Oskar-Morgenstern-Platz 1, 1090 Wien, Austria}

\begin{abstract} 
\noindent
We construct a tensor network representation of the 3d toric code ground state that is stable to a generating set of uniform local tensor perturbations, including those that do not map to local operators on the physical Hilbert space. The stability is established by mapping the phase diagram of the perturbed tensor network to that of the 3d Ising gauge theory, which has a non-zero finite temperature transition. 
More generally, we find that the stability of a topological tensor network state is determined by the form of its virtual symmetries and the topological excitations created by virtual operators that break those symmetries. In particular, a dual representation of the 3d toric code ground state, as well as representations of the X-cube and cubic code ground states, for which point-like excitations are created by such operators, are found to be unstable.
\end{abstract}

\maketitle

\section{Introduction}

\noindent
Tensor network states provide a comprehensive framework for the analytic and numerical study of strongly correlated many-body systems. 
In recent years, this framework has been successfully applied to topological phases of matter~\cite{wegner1971duality,PhysRevLett.50.1395,Wen:1989iv,PhysRevB.82.155138}. For instance, matrix product states (MPS)~\cite{PhysRevLett.69.2863, perezgarcia2006matrix} with projective \emph{virtual} symmetries, which act on the entanglement degrees of freedom, have been utilised to classify one-dimensional symmetry-protected topological phases in terms of their fractionalized boundary modes ~\cite{PhysRevB.84.165139, PhysRevB.83.035107}, while projected entangled-pair states (PEPS)~\cite{verstraete2004renormalization} with matrix product operator virtual symmetries were shown to encode intrinsic topological orders and their anyonic excitations~\cite{SCHUCH20102153,transfermatrix,BUERSCHAPER2014447,csahinouglu2014characterizing,BULTINCK2017183,Bultinck_2017,Williamson:2017uzx}.

From a numerical standpoint MPS perform exceptionally well, underlying the famous density matrix renormalization group algorithm~\cite{PhysRevLett.69.2863}. 
In contrast, the use of PEPS as an ansatz for 2d topological phases is marred by the instability of the topological order to arbitrary perturbations of the tensors~\cite{chen2010tensor,PhysRevB.90.245116,shukla2016boson,garre2017symmetry}. 
This instability is somewhat counter-intuitive since gapped topological phases are stable under local perturbations of the Hamiltonian~\cite{bravyi2010topological}. But local perturbations to the tensors that break the virtual symmetry correspond to non-trivial superselection sectors, and hence cannot be mapped to local physical operators. As a matter of fact, a uniform distribution of such perturbations induces a condensation of the corresponding type of excitations, driving the system into the trivial phase \cite{PhysRevB.79.045316,doi:10.1146/annurev-conmatphys-033117-054154}. 
This is in contrast to virtual symmetry-respecting perturbations which require finite non-zero strength to drive such a phase transition~\cite{shadows,Liu2015,PhysRevB.95.235119}.

This instability against deviations from symmetric tensors is particularly
problematic in numerical simulations, as it implies that one is aiming for
a zero-measure set, and thus variational methods can realize topological
order at best approximately~\cite{PhysRevB.90.245116}. Moreover, to obtain a faithful
approximation of such a zero-measure set, the choice of suitable initial
conditions is typically important. Alternatively, one can restrict the
simulation to symmetric tensors~\cite{he:modular-matrices-tensorRG,iqbal:anyon-orderpars-tc-field}, or try to a posteriori 
extract the symmetry from the optimized tensors~\cite{crone:peps-tcode-detectZ2,francuz:topo-order-from-iPEPS}. These
approaches therefore require a suitable initial guess of the type of
topological order, and potentially carry the risk of biasing the system
through the initial conditions or the symmetries imposed.
Thus, intrinsically stable tensor network representations are 
particularly desirable for the purpose of simulation.

In this work, we construct a tensor network representation for the 3d toric code that is stable under a generating set of uniform local perturbations to the tensors, including those that break the virtual symmetry. 
We prove this statement by relating the quantum phase diagram of the perturbed tensor network to the finite temperature phase diagram of a classical spin system---this is similar to reversing the mapping used in Ref.~\cite{GarciaVerstraeteWolfCirac08}. More specifically, we map the norm of the perturbed tensor network to the quantum partition function of the plaquette components of the toric code Hamiltonian, which is proportional to the partition function of the classical 3d Ising gauge theory~\cite{PhysRevB.78.155120,Li2019}. The stability of our representation then follows from the fact that this model has a non-zero finite temperature phase transition. This strategy is very general and can be applied to various tensor network representations of all Calderbank-Shor-Steane (CSS) stabilizer codes~\cite{PhysRevA.54.1098,Steane2551}. Recent results regarding the stability of these models at finite temperature~\cite{Weinstein2018,Li2019,Weinstein2019} allow us to conjecture a relation between stability and virtual symmetries for a given tensor network representation. 
This provides promising evidence that 3d topological tensor networks form a set of positive measure and hence certain 3d topological phases can be simulated via tensor networks without the introduction of bias or fine-tuning.

\section{Tensor network representations of the 3d toric code} 

\noindent
Let us consider the three-torus $\mathbb T^3$ equipped with a cubic cellulation $\cubu$, whose cubes, plaquettes, edges and vertices are denoted  by $\mathsf c$, $\mathsf{p}$, $\mathsf e$ and $\mathsf v$, respectively. Qubit degrees of freedom are assigned to the edges $\mathsf{e} \subset \cubu$ and governed by the lattice Hamiltonian~\cite{KITAEV20032,PhysRevB.72.035307}  
\begin{equation}
    \label{eq:TC_Ham}
	\mathbb H_{\rm TC}[\cubu] := - \sum_{\mathsf{v}}
	\prod_{\mathsf{e} \supset \mathsf v} Z_{\mathsf e} - \sum_{\mathsf{p}} \prod_{\mathsf{e} \subset \partial \mathsf p} X_{\mathsf e} \, ,
\end{equation}
where $Z$ and $X$ are the standard Pauli matrices. We are interested in (exact) tensor network representations of the ground state subspace of this model. We shall distinguish two representations that are characterised  by different virtual symmetry conditions (see also Ref.~\cite{delcamp2020tensor}).  The first representation we consider is provided  by the tensor network with unit cell
\begin{align}
    \label{eq:unitCellTCOne}
    \includegraphics[scale=1, valign=c]{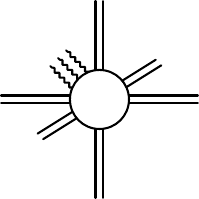} \; := 
    \includegraphics[scale=1, valign=c]{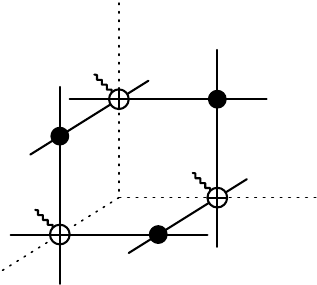} \, ,
\end{align}
where we introduced the tensors
\begin{align*}
    \includegraphics[scale=1, valign=c]{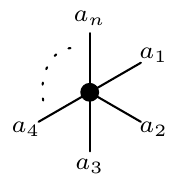} \!\!\! := 
    \prod_{i=1}^{n-1}\delta(a_i+a_{i+1}) \; , \!\!
    \includegraphics[scale=1, valign=c]{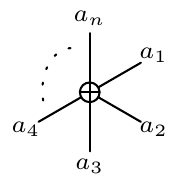} \!\!\! :=  \delta \Big( \sum_{i=1}^n a_i \Big) \, ,
\end{align*}
referred to as $\delta$ and $\delta^\oplus$ tensors, respectively, where all addition is modulo 2.    
In Eq.~\eqref{eq:unitCellTCOne}, solid straight lines correspond to \emph{virtual} indices, while squiggly lines correspond to \emph{physical} indices (this is a commonly used tensor network notation, see Ref.~\cite{TNReview} for a review, but differs from that used in Ref.~\cite{delcamp2020tensor}).  The dotted lines depict edges of the cubic cellulation $\cubu$,  and are included for reference. 
The tensor network state specified by placing copies of the tensor \eqref{eq:unitCellTCOne} on every unit cell of $\cubu$ and contracting the coincident virtual indices of neighboring tensors is described by the equation
\begin{align}
    \label{eq:repTCOne}
	| \psi \ra = {\rm tr}\Big(\bigotimes_{\mathsf e} \mc T_\mathsf{e} \bigotimes_{\mathsf p}\mc T_{\mathsf p}\Big) \, ,
\end{align}
where the local tensors are
\begin{align}
\begin{split}
	\mc T_{\mathsf p} &:= 
	\Big( \prod_{\mathsf e , \mathsf e' \subset  \partial \mathsf p }\delta(a_{\mathsf e , \mathsf p } + a_{\mathsf e',\mathsf p}) \Big)
	\, | \aepSet \ra \, ,
	\\
	\mc T_{\mathsf e} &:= 
	\delta \Big( b_{\mathsf e} + \sum_{\mathsf p \supset \mathsf e}\aep \Big) | b_{\mathsf e} \ra \la \apeSet | \, ,
\end{split}
\end{align}
such that repeated indices are implicitly summed over. 
In the formulae above, $\{b_{\mathsf e}=0,1\}_\mathsf{e}$ correspond to the physical indices, whereas $\{\aep = 0,1\}_{\mathsf p , \mathsf e \subset \partial \mathsf p}$ are the virtual indices.  It follows from the definitions of the $\delta$ and $\delta^\oplus$ tensors that contracting together multiple $\delta$ tensors produces a single $\delta$ with the appropriate number of legs, and similarly for $\delta^{\oplus}$, while 
\begin{align*}
    \includegraphics[scale=1, valign=c]{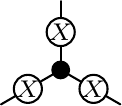} = \includegraphics[scale=1, valign=c]{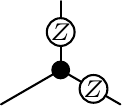} = \includegraphics[scale=1, valign=c]{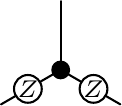} = \frac{1}{2^\frac{1}{2}} \includegraphics[scale=1, valign=c]{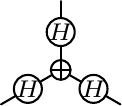} = \includegraphics[scale=1, valign=c]{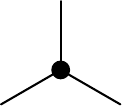}  ,
\end{align*}
with $H$ the \emph{Hadamard} matrix. These defining properties imply that the tensor network \eqref{eq:repTCOne} indeed defines a ground state of the toric code model and hence has topological order~\cite{KITAEV20032,PhysRevB.72.035307,PhysRevB.78.155120}. 
Additionally, we find that the tensor network remains invariant under the action of $X$ operators on the virtual indices along a closed loop in the lattice, where $X$ operators may act on any of the virtual indices adjacent to a lattice edge due to the symmetries of $\delta^\oplus$.
This symmetry is freely deformable and hence is of a topological nature. Since it is codimension-2, it is known as a \text{2-form} symmetry~\cite{Gaiotto:2014kfa}, and as such, we refer to this tensor network as the \emph{2-form} representation.

The second representation is obtained by contracting local tensors associated with the edges and the vertices of $\cubu$ such that a unit cell reads
\begin{equation}
    \label{eq:unitCellTwo}
    \includegraphics[scale=1, valign=c]{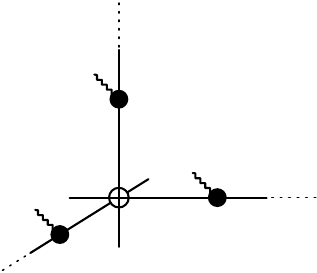} \propto \includegraphics[scale=1, valign=c]{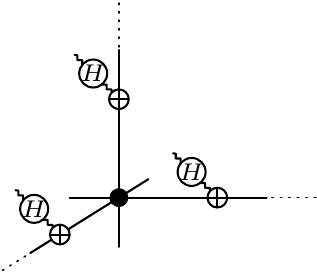}  \, ,
\end{equation}
where the nomenclature is the same as before. The equation above relies on the duality relation between the $\delta$ and $\delta^\oplus$ tensors. In contrast to the \repA representation, this tensor network has a symmetry generated by products of $Z$ operators on the virtual indices that intersect the surface of a cube in the dual lattice, i.e. a deformable codimension-1 virtual symmetry. As such we refer to Eq.~\eqref{eq:unitCellTwo} as the \emph{\repB} representation.

\subsection{Tensor and physical perturbations} 

\noindent
We now consider adding perturbations to the tensor network  and ask whether the topological order is stable under such perturbations. Let us first focus on the \repB representation, which was last introduced. A preliminary question is, which local perturbations of the tensors can be mapped to local physical perturbations? Any virtual perturbation generated by $Z$ operators can be moved locally to the physical indices. But the tensor network is stable under such physical perturbations since the toric code lies in a non-trivial topological phase~\cite{bravyi2010topological}. What about perturbations generated by $X$ operators, which break the virtual symmetry? 
A single virtual $X$ operator corresponds to a point-like topological charge, which cannot be mapped to a local physical operation as it lies in a nontrivial su\-per\-se\-lec\-tion sector~\cite{chen2010tensor}. Given periodic boundary conditions, the relation satisfied by the product of the virtual symmetries over every dual cube imposes that the number of virtual $X$ operators must be even.  Therefore, only a pair of such operators can be mapped to a string-like physical operator. 
As a matter of fact, a uniform $X$ perturbation causes the point-like topological charges to proliferate and condense, driving the system into a trivial phase~\cite{shukla2016boson}. The 1-form representation is thus expected to be \emph{unstable}, which we confirm below.

Repeating the above analysis for the \repA representation, we find that the tensor network is stable under any perturbation generated by $X$ operators as they can be lifted to the physical level. What about $Z$ perturbations? The local relation satisfied by the product of virtual symmetries over the faces of each cube imposes that virtual $Z$ operators can only be inserted in closed (dual) loops, which correspond to topologically nontrivial fluxes, otherwise the tensor network evaluates to zero. It follows that a single virtual $Z$ operator simply annihilates the state, while a closed loop of virtual $Z$ operators maps to a membrane-like physical operator that creates a loop-like flux \cite{PhysRevB.72.035307}. Due to the topology of these excitations, the condensation mechanism that immediately took place in the 1-form representation does not apply here, it is instead exponentially suppressed in the length of the loops. It is thus a priori unclear whether the tensor network is stable under such $Z$ perturbations, and a fortiori under any (uniform) perturbations. 
We focus below on $Z$ perturbations as they generate all topological charge sectors under the virtual symmetries, hence stability to such perturbations is indicative of stability to general local perturbations to the tensors. 

\subsection{Classical partition function and stability under perturbations} 
\noindent
In order to determine the stability of the \repA representation, we shall demonstrate that the norm of the perturbed tensor network can be mapped to the partition function of a classical spin system, such that the quantum phase diagram of the perturbed tensor network reduces to the finite temperate phase diagram of the classical model. Let us consider the following uniform $Z$ perturbation of all the $\mc T_\mathsf{p}$ tensors:
\begin{align}
\label{eq:2FZP}
    \mc T_\mathsf{p} \mapsto \mu \mc T_\mathsf{p} + \nu \widetilde{\mc T}_\mathsf{p}
    \, , 
\end{align}
where $\widetilde{\mc T}_\mathsf{p}$ denotes $\mc T_\mathsf{p}$ multiplied by a single $Z$ operator and $\mu,\nu \in \mathbb C$.
Denoting the perturbed tensor network state by $\ket{\mu,\nu}$, we are interested in the norm of this state obtained by contracting two copies of the tensor network along its physical indices. Using basic properties of the $\delta$ and $\delta^\oplus$ tensors mentioned above, together with the following formula
\begin{equation}
    \label{eq:specProp}
    \includegraphics[scale=1, valign=c]{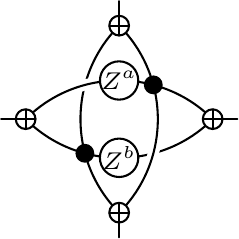} = \delta(a+b) \; \includegraphics[scale=1, valign=c]{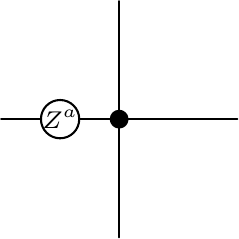} \, ,
\end{equation}
the norm of the perturbed tensor network state reads
\begin{align*}
    \la \mu,\nu | \mu , \nu\ra \propto \bigotimes_\mathsf{e} \big( \la 0 |\mc T_\mathsf{e}\big) \; \bigotimes_\mathsf{p}
    \big(|\mu|^2 \mc T_\mathsf{p} + |\nu|^2 \widetilde{\mc T}_\mathsf{p} \big) \, .
\end{align*}
By shifting the overall normalization of the state by a factor $\nicefrac{1}{\sqrt{\mu}\sqrt{\nu}}$ for each $\mathsf p \subset \cubu$, we can consider instead the norm of the state $| \tilde \mu ,\nicefrac{1}{\tilde \mu} \ra \equiv |\nicefrac{\sqrt{\mu}}{\sqrt{\nu}}, \nicefrac{\sqrt{\nu}}{\sqrt{\mu}}\ra$. Identifying $|\tilde \mu |^2 = e^\beta$, we obtain
\begin{align*}
    \Big\la \tilde \mu , \frac{1}{\tilde \mu} \Big| \tilde \mu , \frac{1}{\tilde \mu} \Big\ra \!
    & \propto 
    \!\!\! \sum_{\{a=0,1\}} \! 
    \prod_{\mathsf e}
    \delta \Big( \sum_{\mathsf p \supset \mathsf e}  a_{\mathsf p} \Big)
    \prod_{\mathsf p}
    (e^\beta + (-1)^{a_\mathsf{p}}e^{-\beta}) \, .
\end{align*}
Utilising the Fourier transform
\begin{align}
    \delta \Big( \sum_{\mathsf p \supset \mathsf e}  a_{\mathsf p}\Big) = \frac{1}{2}\sum_{\sigma_{\mathsf e}=\pm 1}\prod_{ \mathsf p \supset \mathsf e}
    \sigma_{\mathsf e}^{a_\mathsf p} \, ,
\end{align}
we finally obtain
\begin{align}
    \nn
    \Big\la \tilde \mu , \frac{1}{\tilde \mu} \Big| \tilde \mu , \frac{1}{\tilde \mu} \Big\ra \! 
    & \propto \!\!\! \sum_{\{\sigma=\pm 1\}} \!
    \prod_{\mathsf p}\Big( \sum_{a_{\mathsf p}}(e^\beta + (-1)^{a_\mathsf p}e^{-\beta})
    \prod_{\mathsf e \subset \partial \mathsf p}\sigma_{\mathsf e}^{a_\mathsf p}\Big)
    \\ \nn
    & \propto \!\!\! \sum_{\{\sigma = \pm 1 \}} \! \prod_\mathsf p {\rm exp}\Big( \beta \prod_{\mathsf e \subset \partial \mathsf p } \sigma_\mathsf e\Big) = \mc Z_{\rm gauge}[\beta] \, ,
\end{align}
which we recognize as the partition function of the classical 3d Ising gauge theory~\cite{wegner1971duality} at inverse temperature $\beta = 2 \log |\tilde \mu | $. 
The proportionality constant above is simply a power of $2$ that does not appear in physical expectation values and hence we need not keep track of it. 
Crucially, a loop of physical $X$ operators for the quantum model, which serves as a generalized order parameter for the $Z$ perturbation driven quantum phase transition~\cite{Gaiotto:2014kfa}, is directly mapped to a \emph{Wilson} loop operator for the statistical model, i.e. 
\begin{align}
\label{eq:Xoperatormapping}
    \Big\la \tilde \mu , \frac{1}{\tilde \mu} \Big| \prod_{\mathsf e \subset \mc C }X_\mathsf e \Big| \tilde \mu , \frac{1}{\tilde \mu} \Big\ra
    \Big / \Big\la \tilde \mu , \frac{1}{\tilde \mu} \Big| \tilde \mu , \frac{1}{\tilde \mu} \Big\ra
    &= 
    \Big\langle \prod_{\mathsf e\subset \mc C}\sigma_\mathsf e \Big\rangle_{\! \beta}  \, ,
\end{align}
where $\mc C$ denotes a closed planar loop along the edges of the lattice.  
It follows that a finite temperature phase transition, measured by the loop order parameter above, occurs in the classical model if and only if a phase transition occurs in the quantum tensor network, w.r.t. the $X$ loop order parameter above.
Since the 3d Ising gauge theory is known to have a finite temperature phase transition, the quantum tensor network has a phase transition at finite perturbation strength. 
Moreover, the $\beta \to \infty$ limit of the partition function corresponds to the norm of the unperturbed tensor network and can be shown to reduce to the partition function of $\mathbb Z_2$-BF theory \cite{Horowitz:1989ng}, i.e.
\begin{align}
    \mc Z_{\rm gauge}[\infty] = \sum_{\{\sigma =\pm 1\}} \!
    \prod_\mathsf p \delta \Big(\prod_{\mathsf e \subset \partial \mathsf p} \sigma_\mathsf e\Big) \, ,
\end{align}
where the expectation value of the loop operator is one. Therefore, for sufficiently small perturbation strength $|\nu|\ll |\mu|$ (or $|\tilde \mu| \gg 1 $), the resulting 3d Ising gauge theory lies in the deconfined/topological phase, where the Wilson loop operator satisfies a perimeter law~\cite{wegner1971duality,Sachdev_2018}, and thus the representation is \emph{stable}. As we increase the perturbation strength, the system undergoes a second order phase transition towards the confining phase of the gauge theory, where the Wilson loop operator satisfies an area law due to an increase in fluctuations of the plaquette fluxes~\cite{wegner1971duality,Sachdev_2018}, at which point the representation is no longer stable. Alternatively, this phase transition can be described as spontaneous \emph{higher-form} symmetry breaking~\cite{Gaiotto:2014kfa}, with respect to the 1-form symmetry generated by the vertex terms $\prod_{\mathsf e \supset \mathsf v}Z_\mathsf e$, such that the symmetry broken ordered phase corresponds to the topological deconfined one. Similarly, given that $Y=iXZ$, we can show that this representation is stable to a uniform $Y$ perturbation as the corresponding classical partition function is the same as for a $Z$ perturbation. Indeed, in the computation of the norm, the $Y$ operators must appear at the same position in bra and ket layers for a non-zero overlap, in which case a pair $X$ operators can be moved to the physical level and cancelled. 
Stability to general single site perturbations is discussed App.~\ref{sec:app_topoTNCSS}.

Following a similar approach, we can confirm that the other basis is unstable because the norm of the perturbed tensor network maps to a classical statistical model with a zero temperature phase transition. Using the alternative form of the tensor network given on the r.h.s of \eqref{eq:unitCellTwo} and mimicking the previous computation, we find that the norm of the tensor network for a perturbation of the vertex tensors reads
\begin{align}
    \label{eq:classicModelOneForm}
    \Big\la \tilde \mu , \frac{1}{\tilde \mu} \Big| \tilde \mu , \frac{1}{\tilde \mu} \Big\ra \! 
    & \propto \!\!\! 
    \sum_{\{\sigma = \pm 1 \}} \! \prod_\mathsf v {\rm exp}\Big( \beta \prod_{\mathsf e \supset \mathsf v} \sigma_\mathsf e\Big) \, ,
\end{align}
for which (see App.~\ref{sec:app_oneForm})
\begin{equation}
    \label{eq:expansionWilson}
     \Big\langle \prod_{\mathsf e\subset \partial \mc R}\sigma_\mathsf e \Big\rangle_{\! \beta} 
     \sim 
     {\rm tanh}(\beta)^{|\mathcal R|}
     \, ,
\end{equation}
where $\partial \mc R$ is a closed surface along the plaquettes of the dual lattice, and $|\mc R|$ denotes the number of vertices enclosed within $\mc R$. Since the generalized order parameter obeys a volume law for any finite $\beta$, the classical model has a zero temperature phase transition, hence the instability of the \repB representation.

\section{Tensor network representations and stability of the X-cube model\label{sec:app_XCube}} 

\noindent
In this section, we study the stability of the tensor network representations of the X-cube model, which is a type-I fracton model. Qubit degrees of freedom are assigned to the edges of $\cubu$ governed by the Hamiltonian \cite{PhysRevB.94.235157}
\begin{align}
\label{eq:XcubeH}
    \mathbb H_{\rm X}[\cubu] = -\sum_{\mathsf v}\Big( \prod_{\substack{\mathsf e \supset \mathsf v \\ \mathsf e \perp \hat{x}}} Z_\mathsf e  +  \prod_{\substack{\mathsf e \supset \mathsf v \\  \mathsf e \perp \hat{y}}} Z_\mathsf e \Big) - \sum_{\mathsf c} \prod_{\mathsf e \in \partial \mathsf c} X_\mathsf e 
    \, ,
\end{align}
where $\hat{x}$ and $\hat{y}$ are two orthogonal vectors that go along the edges of $\cubu$. Applying the same approach as for the toric code model, we find two tensors network representations of the ground state sector. The first representation is in terms of local tensors associated with the edges and the cubes of $\cubu$, whereas the second one is in terms of pairs of local tensors associated with the vertices and tensors associated with the edges. Using the same notation as above, the unit cells of these representations are 
\begin{align}
    \label{eq:unitCellX}
    \!\!\! \includegraphics[scale=1, valign=c]{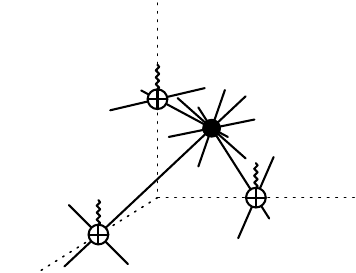} \;\; {\rm and} \!\!  \includegraphics[scale=1, valign=c]{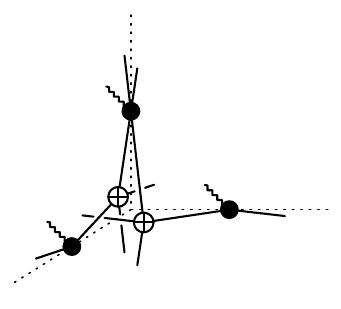} , 
\end{align}
respectively, where the $\delta^\oplus$ tensors on the r.h.s are both associated with the same vertex of $\cubu$. 

Let us focus for now on the representation depicted on the l.h.s of \eqref{eq:unitCellX}. The tensor network state is stable under perturbations of the $\delta^\oplus$ tensors generated by $X$ operators, since these can be lifted to physical indices. What about $Z$ perturbations? Since the network remains invariant under the action of $X$ operators that form `cages' along the virtual indices~\cite{prem2018cage}, implied by the $X$-stabilizers in Eq.~\eqref{eq:XcubeH}, the number of $Z$ operators must be even on all dual lattice planes, otherwise the tensor network evaluates to zero. This implies that a given $Z$ perturbation, which corresponds to a \emph{fracton} excitation~\cite{PhysRevB.94.235157}, cannot be mapped to a local physical operator. Similar to the \repB representation of the toric code, a uniform $Z$ perturbation induces a condensation of the fractons causing the tensor network to enter the trivial phase, even for arbitrarily small perturbation strength. These can be confirmed by computing the norm of the perturbed tensor:
\begin{align}
    \Big\la \tilde \mu , \frac{1}{\tilde \mu} \Big| \tilde \mu , \frac{1}{\tilde \mu} \Big\ra 
    \propto
    \sum_{\{\sigma=\pm 1\}}
    \prod_{\mathsf c}{\rm exp}\Big( \beta \prod_{\mathsf e \subset \partial \mathsf c} \sigma_{\mathsf e}\Big)
    \, ,
\end{align}
along with the fact that $X$ operators are mapped to classical spin operators.  
We identify this norm as the quantum partition function associated with the $X$ stabilizers of the X-cube model (see App.~\ref{sec:app_ft}). It follows from the analysis carried out in~\cite{Weinstein2018,Li2019,Weinstein2019} that the resulting classical spin model has a zero temperature phase transition to the trivial phase, consistent with the phase transition of the tensor network due to fracton condensation. 

Similarly, the representation depicted on the r.h.s. of \eqref{eq:unitCellX} is stable under perturbations of the $\delta$ tensors generated by $Z$ operators. Moreover, for every plane of $\cubu$, the network has a virtual symmetry generated by products of $Z$ operators along closed loops of the dual 2d lattice associated with this plane. This implies that the number of $X$ perturbations on every plane must be even, so that a single $X$ operator, which correspond to a \emph{lineon} excitation~\cite{PhysRevB.94.235157}, cannot be lifted locally to a physical operator. A uniform $Z$ perturbation thus induces a condensation of the lineons, making this representation unstable. Correspondingly, the norm of the perturbed tensor network again maps to a 3d classical generalized Ising model~\cite{PhysRevB.94.235157} (with $Z$ operators mapping to classical spin operators)
\begin{align}
    \Big\la \tilde \mu , \frac{1}{\tilde \mu} \Big| \tilde \mu , \frac{1}{\tilde \mu} \Big\ra
    \! \propto \!\!\!
    \sum_{\{\sigma = \pm 1\}} \!
    \prod_\mathsf v
    {\rm exp} \Big( \beta
    \prod_{\substack{\mathsf e \supset \mathsf v \\ \mathsf e \perp \hat{y}}} \sigma_{\mathsf e} + \beta \prod_{\substack{\mathsf e \supset \mathsf v \\ \mathsf e \perp \hat{y}}} \sigma_{\mathsf e} \Big)
    \, , 
\end{align}
which has a zero temperature phase transition to the trivial phase~\cite{Weinstein2018,Li2019,Weinstein2019}. In sharp contrast with the 3d toric code model, both representations turn out to be unstable for the X-cube model.

\section{Tensor network representations and stability of Haah's cubic code\label{sec:app_HaahCode}} 

\noindent
In this section, we study the stability of the tensor network representations of Haah's cubic code, which is a type-II fracton model. Pairs of qubit degrees of freedom are assigned to the vertices of $\cubu$ governed by the lattice Hamiltonian~\cite{PhysRevA.83.042330}
\begin{align}
    \mathbb H_{\rm Haah}[\cubu] = -\sum_{\mathsf c} \mathbb X(\mathsf c) -\sum_{\mathsf c} \mathbb Z(\mathsf c) \, ,
\end{align}
such that
\begin{align}
    \mathbb X (\mathsf c) &= 
    \prod_{\mathsf v \in N^+_\mathsf c}(IX)_\mathsf v 
    \prod_{\mathsf v \in \widetilde{N}^+_\mathsf c}(XI)_\mathsf v
    \\
    \mathbb Z (\mathsf c) &= 
    \prod_{\mathsf v \in N^-_\mathsf c}(ZI)_\mathsf v 
    \prod_{\mathsf v \in \widetilde{N}^-_\mathsf c}(IZ)_\mathsf v \, ,
\end{align}
where $N^\pm_\mathsf c = \{\mathsf v^\pm_\mathsf c, \mathsf v^\pm_\mathsf c \pm \hat x, \mathsf v^\pm_\mathsf c \pm \hat y , \mathsf v^\pm_\mathsf c  \pm \hat z\}$ and $\widetilde{N}^\pm_\mathsf c = \{\mathsf v^\pm_\mathsf c , \mathsf v^\pm_\mathsf c \pm \hat x \pm \hat y, \mathsf v^\pm_\mathsf c \pm \hat y \pm \hat z , \mathsf v^\pm_\mathsf c  \pm \hat x \pm \hat z\}$ for $\mathsf v^+_\mathsf c$ ($\mathsf v^-_\mathsf c$) the corner of $\mathsf c$ with minimal (maximal) $(x,y,z)$ coordinates. 
Since the $X$ and $Z$ stabilizers of cubic code are related by a duality, we only need consider the tensor network representation of the ground state sector obtained by contracting the following local tensors:
\begin{align*}
    \mc T_{\mathsf v} &:= 
	\delta \Big( b_{\mathsf v} + \sum_{\mathsf c \in N_{\mathsf v}}a_{\mathsf v, \mathsf c} \Big) | b_{\mathsf v} \ra \la \{a_{\mathsf v , \mathsf c}\}_{\mathsf c \in N_\mathsf v} | \, ,
    \\
    \widetilde{\mc T}_{\mathsf v} &:= 
	\delta \Big( \widetilde{b}_{\mathsf v} + \sum_{\mathsf c \in \widetilde{N}_{\mathsf v}}\widetilde{a}_{\mathsf v, \mathsf c} \Big) | \widetilde{b}_{\mathsf v} \ra \la \{\widetilde{a}_{\mathsf v , \mathsf c}\}_{\mathsf c \in \widetilde{N}_\mathsf v} | \, ,
    \\
	\mc T_{\mathsf c} &:= 
	\Big( \!\!\! \prod_{\mathsf v , \mathsf v' \in N^+_\mathsf c \cup \widetilde{N}^+_\mathsf c} \!\!\!\!\!\! \delta(a_{\mathsf v , \mathsf c } + a_{\mathsf v',\mathsf c}) \Big)
	\, | \{a_{\mathsf v, \mathsf c}\}_{\mathsf v \in N^+_\mathsf c}, \{\widetilde{a}_{\mathsf v, \mathsf c}\}_{\mathsf v \in \widetilde{N}^+_\mathsf c} \ra \, ,
\end{align*}
such that $N_\mathsf v=\{\mathsf  c_\mathsf v,\mathsf c_\mathsf v -\hat{x},\mathsf c_\mathsf v -\hat{y},\mathsf c_\mathsf v -\hat{z}\}$ and $\widetilde{N}_\mathsf v=\{\mathsf c_\mathsf v ,\mathsf c_\mathsf v -\hat{x}-\hat{y},\mathsf c_\mathsf v -\hat{y}-\hat{z},\mathsf c_\mathsf v -\hat{x}-\hat{z}\}$ for $\mathsf c_\mathsf v$ the cube at coordinate $\mathsf v+\frac{1}{2}(\hat{x}+\hat{y}+\hat{z})$. This tensor network has unit cell 
\begin{equation}
	\includegraphics[scale=1, valign=c]{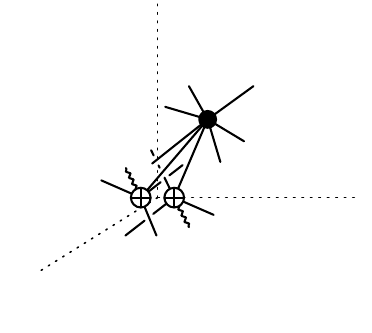}
\end{equation}
and virtual symmetries that are obtained from the $X$-stabilizers of the cubic code. 
On periodic boundary conditions, these virtual symmetries lead to a number of fractal-like global relations (products that yield the identity), whose exact number sensitively depends on the precise system size, but is bounded by $e^{c L}$ for the linear extent $L$ and a constant $c$. 
The number of virtual $Z$ operators must be even over the tensors involved in all such relations, leading to the immobility of a fracton excitation.

Virtual $X$ perturbations map to physical operators, while $Z$ perturbations induce a condensation of fractons that instantly drives a phase transition to the trivial phase. 
The norm of the perturbed tensor network maps to a 3d classical fractal Ising model~\cite{PhysRevB.94.235157,PhysRevBUngauging} (along with $X$ operators mapping to classical spin operators) 
\begin{align}
    \Big\la \tilde \mu , \frac{1}{\tilde \mu} \Big| \tilde \mu , \frac{1}{\tilde \mu} \Big\ra  = 
    \!\!\! \sum_{\{\sigma, \sigma'= \pm 1\}} \!
    \prod_\mathsf c 
    {\rm exp} \Big( \beta
    \prod_{\mathsf v  \in {N}^+_\mathsf c} \sigma_{\mathsf v}  \prod_{\mathsf v \in \widetilde{N}^+_\mathsf c} \sigma_{\mathsf v}' \Big)
    \, , 
\end{align}
which has a zero temperature phase transition to the trivial phase~\cite{Weinstein2018,Li2019,Weinstein2019}, hence the instability of the tensor network state.

\section{Generalization and discussion} 

\noindent
We have demonstrated that the toric code model admits two tensor network representations, which behave differently under perturbations.  
More generally, any CSS stabilizer code Hamiltonian has the form 
\begin{align}
    \mathbb H_{\text{CSS}} = \lambda_\mathbb A \mathbb H_\mathbb A(Z) + \lambda_\mathbb B \mathbb H_\mathbb B(X) \, ,
\end{align} 
where $\mathbb H_\mathbb A(Z)$ and $\mathbb H_\mathbb B(X)$ are local commuting projector Hamiltonians expressed solely in terms of Pauli $Z$ and $X$ operators, respectively. For such a Hamiltonian we can always define two canonical tensor network representations of its ground state sector. These are found by first enforcing either the $X$ or $Z$ stabilizers,  via a choice of initial product state in the corresponding basis, followed by projection with respect to the remaining stabilizers.  
We can then ask which of these representations---if any---are stable. In 2d all topological stabilizer codes are equivalent to copies of the toric code~\cite{bombin2012universal,bombin2014structure,Haah2018a}, and so we expect them to be unstable. In 3d there is however a wide variety of inequivalent stabilizer codes due to the existence of fracton topological order~\cite{doi:10.1146/annurev-conmatphys-031218-013604,Pretko2020} (see the appendix of Ref.~\cite{Dua2019a} for a collection of codes). 

Generally, in order to determine the stability of a given tensor network state under arbitrary local perturbations, we exploit the fact that the norm of the perturbed tensor network can be mapped to that of the quantum partition function of the stabilizer Hamiltonian for $\lambda_\mathbb A = 0$ or $\lambda_\mathbb B = 0$, depending on the representation. This partition function can in turn be rewritten as that of a classical spin model, for which the existence of a non-zero finite temperature phase transition translates into the stability of the representation (see App.~\ref{sec:app_topoTNCSS} \& \ref{sec:app_ft}). 
This raises the question, do (un)stable representations share distinctive features? 

Given a classical spin model, a zero-temperature phase transition will occur if the spins are essentially \emph{free}, i.e. the number of relations satisfied by the spins does not grow according to the volume of the lattice. But these relations descend from the constraints that the operators of the underlying stabilizer Hamiltonian obey. At the level of the tensor network, we recover these constraints in the form of redundancies of the virtual symmetries. These redundancies in turn dictate the topology of the \emph{virtual charges}, i.e. operator insertions breaking virtual symmetry conditions that correspond to non-trivial topological excitations on the physical level. As long as the virtual charges are point-like, which is the case for all the 3d tensor network states we considered apart from the 2-form representation of the toric code, the number of independent redundancies can grow at most linearly with the linear extent of the system. This signifies that the spins of the corresponding classical model must fulfil a number of constraints whose number becomes negligible in the thermodynamic limit, hence a zero-temperature phase transition. Conversely, the existence of non-trivial independent local relations whose number scales with the volume of the lattice, which implies the existence of extended virtual charges, should guarantee a non-zero finite temperature phase transition. For instance, the 4d 2-form $\mathbb Z_2$ gauge model, which hosts electric and magnetic loop-like excitations, admits two tensor network representations that turn out to be stable to perturbations. Putting everything together, we conjecture that \emph{a CSS topological tensor network state that supports strictly extended virtual charges is stable under arbitrary (infinitesimal) perturbations.} 
This would immediately imply that the ground spaces of all 2d and 3d topological stabilizer Hamiltonians admit an unstable tensor network representation as they are known to support point-like topological charges~\cite{haah2013commuting}.

\bigskip
\noindent\textbf{Acknowledgements:} 
\emph{CD would like to thank Markus Hauru for stimulating discussions about closely related topics. 
DW acknowledges support from the Simons foundation. 
This project has received funding from the European Research Council (ERC) under the European Union’s Horizon 2020 research and innovation programme through the ERC Starting Grant WASCOSYS (No.~636201)  and the ERC Consolidator Grant SEQUAM (No.~863476), as well as the Deutsche Forschungsgemeinschaft (DFG, German Research Foundation) under Germany’s Excellence Strategy -- EXC-2111 -- 390814868.}

\bibliography{Refs}

\clearpage 
\appendix 

\onecolumngrid


\section{1-form representation of the $d$-dimensional toric code\label{sec:app_oneForm}}
\noindent
In this section we show that the 1-form representation of the $d$-dimensional toric code is always unstable. We consider the $d$-torus $\mathbb T^d$ equipped with a $d$-dimensional hypercubic cellulation $\mathbb T^d_\boxempty$. Qubit degrees of freedom are still assigned to edges $\mathsf e \subset \mathbb T^d_\boxempty$ and they are governed by a lattice Hamiltonian of the same form as in 
the main text 
with stabilizer generators acting on vertices and plaquettes, respectively. 

As in 3d, the 1-form representation of the ground state is obtained by contracting $\delta$ and $\delta^\oplus$ tensors associated with the edges and the vertices of $\mathbb T^d_\boxempty$, respectively, according to a pattern that is the obvious generalization of 
that presented in the main text. 
The resulting tensor network state has a virtual symmetry generated by product of $Z$ operators on virtual indices that intersect the surface of a $d$-cube in the dual hypercubic lattice.

Perturbations of the local tensors generated by $Z$ operators can be lifted locally to the physical sector, and as such the tensor network state is stable under such perturbations. Let us now consider a uniform $X$ perturbation of the $\delta$ tensors. Using the duality relation between $\delta$ and $\delta^\oplus$ tensors, and adapting in an obvious way the derivation in the main text, we find that the norm of the perturbed tensor network state reads
\begin{align}
    \label{eq:classicModelOneForm}
    \Big\la \tilde \mu , \frac{1}{\tilde \mu} \Big| \tilde \mu , \frac{1}{\tilde \mu} \Big\ra \! 
    & \propto \!\!\! 
    \sum_{\{\sigma = \pm 1 \}} \! \prod_\mathsf v {\rm exp}\Big( \beta \prod_{\mathsf e \supset \mathsf v} \sigma_\mathsf e\Big) 
\end{align}
such that the expectation value of the relevant generalized order parameter satisfies
\begin{align}
    \Big\la \tilde \mu , \frac{1}{\tilde \mu} \Big| \prod_{\mathsf e \subset \partial \mc R}Z_\mathsf e \Big| \tilde \mu , \frac{1}{\tilde \mu} \Big\ra
    \Big / \Big\la \tilde \mu , \frac{1}{\tilde \mu} \Big| \tilde \mu , \frac{1}{\tilde \mu} \Big\ra
    &= 
    \Big\langle \prod_{\mathsf e\subset \partial \mc R}\sigma_\mathsf e \Big\rangle_{\! \beta} \, ,
\end{align}
where $\partial \mc R$ is a closed surface along the edges of the dual lattice.
We now would like to show that the resulting classical model has a zero temperature phase transition, which in turn implies that the tensor network state is unstable. To proceed, it is convenient to use an alternative expression for \eqref{eq:classicModelOneForm}, which we obtain once again by analogy with the derivation presented in the main text:
\begin{align}
    \label{eq:altOneFormExp}
    \Big\la \tilde \mu , \frac{1}{\tilde \mu} \Big| \tilde \mu , \frac{1}{\tilde \mu} \Big\ra \!
    & \propto 
    \!\!\! \sum_{\{a=0,1\}} \! 
    \prod_{\mathsf e}
    \delta \Big( \sum_{\mathsf v \subset \partial \mathsf e}  a_{\mathsf v} \Big)
    \prod_{\mathsf v}
    (e^\beta + (-1)^{a_\mathsf{v}}e^{-\beta}) \, ,
\end{align}
where $|\tilde \mu|^2 = e^\beta$. This partition function is easy to evaluate explicitly and we find
\begin{align}
    \label{eq:altOneForm}
    \Big\la \tilde \mu , \frac{1}{\tilde \mu} \Big| \tilde \mu , \frac{1}{\tilde \mu} \Big\ra \!
    & \propto 
    (e^\beta + e^{- \beta})^{|\mathbb T^d_\boxempty|}
    +
    (e^\beta - e^{- \beta})^{|\mathbb T^d_\boxempty|} \, ,
\end{align}
where $|\mathbb T^d_\boxempty|$ equals the number of vertices in $\mathbb T^d_\boxempty$. It is now apparent that the partition function \eqref{eq:classicModelOneForm} can be mapped to that of a 1d classical Ising model, which is known to have a zero temperature phase transition. This can be appreciated by considering the expectation value of the generalized order parameter. Given that all the configuration variables are implicitly identified in \eqref{eq:altOneFormExp}, we should interpret $\la \tilde \mu ,\nicefrac{1}{\tilde \mu} | \prod_{\mathsf e \subset \partial \mc R}Z_\mathsf e  | \tilde \mu ,\nicefrac{1}{\tilde \mu} \ra $ as the partition function of the classical model with a  domain wall that coincides with $\partial \mc R$ so that
\begin{align}
    \nn
    \Big\la \tilde \mu , \frac{1}{\tilde \mu} \Big| \prod_{\mathsf e \subset \partial \mc R}
    Z_\mathsf e \Big| \tilde \mu , \frac{1}{\tilde \mu} \Big\ra
    \Big / \Big\la \tilde \mu , \frac{1}{\tilde \mu} \Big| \tilde \mu , \frac{1}{\tilde \mu} \Big\ra
    &= 
    \frac{(e^\beta - e^{-\beta})^{|\mc R|}(e^\beta + e^{-\beta})^{|\mathbb T^d_\boxempty|-|\mc R|}  + (e^\beta + e^{-\beta})^{|\mc R|}(e^\beta - e^{-\beta})^{|\mathbb T^d_\boxempty|-|\mc R|}}
    {(e^\beta + e^{-\beta})^{|\mathbb T^d_\boxempty|}  + (e^\beta - e^{-\beta})^{|\mathbb T^d_\boxempty|}}
    \\
    \nn
    &=
    \tanh(\beta)^{|\mc R|}
    \frac{1  + \tanh(\beta)^{|\mathbb T^d_\boxempty|-2|\mc R|}}
    {1  + \tanh(\beta)^{|\mathbb T^d_\boxempty|}}
    \\
    &\sim
    \tanh(\beta)^{|\mc R|} 
    \q  \text{for} \;  |\mc R| \ll |\mathbb T^d_\boxempty| \, ,
\end{align}
where $| \mc R|$ equals the number of vertices in $\mathbb T^d_\boxempty$ that are enclosed by $\partial \mc R$. The expectation value is exponentially decaying with the volume of the region $\mc R$ for any finite value of $\beta$, and is equal to 1 at $\beta \to \infty$. This confirms that any non-trivial uniform $X$ perturbation induces a phase transition, and as such the 1-form representation of the toric code model is always unstable.

\section{Topological stabilizer tensor networks\label{sec:app_topoTNCSS}}

\noindent
In this section, we explain how, given a CSS stabilizer code, we can construct a canonical tensor network representation of its ground state subspace. Furthermore, we describe how to construct an isometric tensor associated with a region of the network and specify how topological excitations can be encoded within this framework in terms of virtual operators.

\subsection{Definition of the tensor network}

\noindent
Before specializing to CSS stabilizer codes, let us consider a model with microscopic degrees of freedom associated to sites $\mathsf s \in \mc S$ governed by a general frustration-free local commuting projector Hamiltonian
\begin{align}
    \mathbb H = - \sum_{\mathsf i \in \mc I} \mathbb P_\mathsf i 
    \q \q {\rm with} \q 
    \mathbb P_\mathsf i^\dagger = \mathbb P_\mathsf i
    \q {\rm and} \q
    [\mathbb P_\mathsf i,\mathbb P_{\mathsf i'}]=0 \;\; \forall \, \mathsf i,\mathsf i' \in \mc I
    \, ,
\end{align}
where $\mc I$ denotes the set of \emph{interactions}. Given that all the local projectors commute with one another, this Hamiltonian admits a ground state projector $\prod_{\mathsf i \in \mc I} \mathbb P_\mathsf i$, which can be naturally interpreted as a tensor network. In order to derive such a tensor network, we consider a tensor rank decomposition of each projector
\begin{align}
    \mathbb P_\mathsf i = \sum_{k=1}^{{\rm rank}(\mathbb P_\mathsf i)} 
    m^{(k)} \bigotimes_{\mathsf s \in N_\mathsf i} \mathbb M_\mathsf s^{(k)} \, ,
\end{align}
where $N_\mathsf i$ denotes the set of \emph{sites} $\mathsf s$ acted upon by $\mathbb P_\mathsf i$, and $\mathbb M_\mathsf s$ are single site operators. This decomposition can be interpreted as the following tensor network obtained by contracting local tensors associated with the interaction $\mathsf i$ and sites $\mathsf s \in N_\mathsf i$:
\begin{align}
    \mathbb P_\mathsf i = {\rm tr}\Big(\mc T_\mathsf i \bigotimes_{\mathsf s \in N_\mathsf i}\mc T_\mathsf s^\mathsf i \Big)
    \q \q {\rm with} \q 
    \mc T_\mathsf s^\mathsf i = \mathbb M_\mathsf s^{(k_{\mathsf s,\mathsf i})} \otimes \bra{k_{\mathsf s, \mathsf i}}
    \q {\rm and} \q
    \mc T_\mathsf i = \Big(\prod_{\mathsf s , \mathsf s' \in N_\mathsf i}
    \!\!
    \delta(k_{\mathsf s , \mathsf i}+k_{\mathsf s',\mathsf i}) \Big) m^{(k_{\mathsf s,\mathsf i})} \ket{\{k_{\mathsf s,\mathsf i}\}}_{\mathsf s \in N_{\mathsf i}} 
    \, ,
\end{align}
where repeated $k$-indices are contracted. Contracting the small tensor network given above for each projector $\mathbb P_\mathsf i$ yields a tensor network representation of the ground state projector $\prod_{\mathsf i \in \mc I}\mathbb P_\mathsf i$.
Given a product state $\bigotimes_{\mathsf s \in \mc S}| \psi_\mathsf s \ra$ such that $\langle \prod_{\mathsf i \in \mc I} \mathbb P_\mathsf i \rangle \neq 0$ with respect to $\bigotimes_{\mathsf s \in \mc S} \ket{\psi_\mathsf s}$, we finally obtain an unnormalized tensor network representation for a ground state $| \psi \ra$ by applying the ground state projector to it:
\begin{align}
    \label{eq:CPTN}
    | \psi \ra = \Big( \prod_{\mathsf i \in \mc I} \mathbb P_\mathsf i \Big) \bigotimes_{\mathsf s \in \mc S} \ket{\psi_\mathsf s}
    \, .
\end{align}

\bigskip \noindent
Let us now focus on \emph{Pauli stabilizer models}, for which the physical degrees of freedom are \emph{qubits} and the commuting projectors $\mathbb P_\mathsf i$ can be conveniently expressed in terms of Pauli operators that generate the stabilizer group. For simplicity, we further restrict to a class of stabilizer models known as CSS codes~\cite{PhysRevA.54.1098,Steane2551}. 
In this case, the operators can be partitioned into Pauli-$X$ terms $\{\mathbb B_\mathsf i(X)\}_{\mathsf i \in \mc I^X}$ and Pauli-$Z$ terms $\{\mathbb A_\mathsf i(Z)\}_{\mathsf i \in \mc I^Z}$ such that
\begin{equation}
    \mathbb H = - \sum_{\mathsf i \in \mc I^Z}\mathbb A_\mathsf i(Z) - \sum_{\mathsf i \in \mc I^X}\mathbb B_\mathsf i(X)
    \q\q {\rm with} \q \mathbb A_\mathsf i (Z)= \prod_{\mathsf s \in N_\mathsf i} Z_\mathsf s 
    \q {\rm and} \q
    \mathbb B_\mathsf i (X)= \prod_{\mathsf s \in N_\mathsf i} X_\mathsf s \, ,
\end{equation}
where as before $N_\mathsf i$ denotes the set of sites/qubits acted upon by the corresponding operator. 
Since the projectors $\frac{1}{2}({\rm id}+\mathbb A_\mathsf i(Z))$ act trivially on $|0\ra^{\otimes |\mc S|}$, the ground state admits a particularly simple tensor network representation, namely
\begin{equation}
    | \psi \ra = \Big(\prod_{\mathsf i \in \mc I^X} \frac{1}{2}\big({\rm id}+\mathbb B_\mathsf i (X) \big)\Big)
    | 0\ra^{\otimes |\mc S|} \, ,
\end{equation}
where $|\mc S|$ equals the number of sites/qubits in the system.
We note that in this case $\bra{0}^{\otimes |\mc S|}  \prod_\mathsf i ({\rm id}+ \mathbb B_\mathsf i)  \ket{0}^{\otimes |\mc S|}  \geq 1$ provides the number of $X$ stabilizer \emph{constraints}, i.e., products of distinct $\mathbb B_\mathsf i(X)$ operators that are equal to the identity map (including the empty product).  We further remark that the ${\rm log}_2$ of this number equals the number of independent stabilizer constraints. 
The above tensor network can be written, up to an overall normalization, as
\begin{equation}
    | \psi \ra = {\rm tr}\Big(\bigotimes_{\mathsf s \in \mc S} \mc T_\mathsf s \bigotimes_{\mathsf i \in \mc I^X}\mc T_\mathsf i \Big)
\end{equation}
in terms of the local tensors 
\begin{align}
	\mc T_{\mathsf s} = 
	\delta \Big( b_{\mathsf s} + \sum_{\mathsf i \in N_\mathsf s}
	a_{\mathsf s, \mathsf i}\Big) | b_{\mathsf s} \ra \la \{a_{\mathsf s , \mathsf i}\}_{\mathsf i \in N_\mathsf s} | 
	\q {\rm and} \q
	\mc T_{\mathsf i} = 
	\Big( \prod_{\mathsf s , \mathsf s' \in N_\mathsf i }\delta(a_{\mathsf s , \mathsf i } + a_{\mathsf s',\mathsf i}) \Big)
	\, | \{a_{\mathsf s , \mathsf i}\}_{\mathsf s \in N_\mathsf i}\ra \, ,
\end{align}
such that repeated indices are contracted, where $N_\mathsf s$ denotes the set of $\mathbb B_\mathsf i(X)$ operators that act on the qubit $\mathsf s$ and $\{a,b=0,1\}$ are $\mathbb Z_2$ variables in the Pauli-$Z$ basis. These tensors are contracted according to the bipartite \emph{interaction graph} associated with the $\mathbb B_\mathsf i (X)$ operators, i.e. the graph with vertices for the interactions $\mathsf i$ and qubits $\mathsf s$ that are linked by an edge when the generator $\mathbb B_\mathsf i(X)$ acts non-trivially on the qubit $\mathsf s$. We further define the bipartite \emph{adjacency matrix} of the interaction graph that is given by the \textit{stabilizer map} $\varsigma_X:\mathbb{Z}_2[\mc I^X] \to  \mathbb{Z}_2[\mc S]$ with columns given by the $\mathbb{Z}_2$-vector representation of the $\{\mathbb B_\mathsf i (X)\}_{\mathsf i \in \mc I^X}$ operators~\cite{Gottesman1997}. In this language, the number  of independent constraints amongst the $X$ stabilizer generators is given by the dimension of the vector space ${\rm ker}\, \varsigma_X$.
Furthermore, the tensor network ground state can be conveniently expressed in terms of the $\delta$ and $\delta^\oplus$ tensors introduced in the main text such that we assign a $\delta$ tensor to each interaction vertex and a $\delta^\oplus$ tensor, with an extra physical leg, to each qubit vertex, the contraction pattern being dictated by the interaction graph, e.g. 
\begin{equation*}
    \includegraphics[scale=1, valign=c]{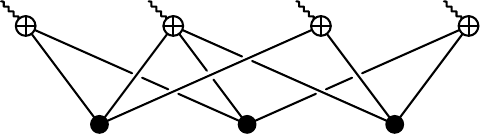} \q {\rm with} \q \varsigma_X = 
    \begin{pmatrix}1 & 1 & 0 \\ 1 & 1 & 1 \\ 1 & 0 & 1 \\ 0 & 1 & 1 \end{pmatrix} \, .
\end{equation*}
The \emph{dual} representation, i.e. the representation obtained by enforcing the $X$ stabilizer constraints initially, can be conveniently obtained by applying the exact same recipe after performing the change of basis $Z \leftrightarrow X$ to the original model.

Operators on the virtual level of the tensor network are generated by products of $X$ and $Z$ operators along the edges of the network. 
It is convenient to move all $X$ operators onto the qubit vertices, and all $Z$ operators onto the interaction vertices. 
The tensor product of $X$ operators on an even number of legs of a qubit vertex tensor $\mc T_\mathsf s$, or $Z$ operators on an even number of legs of an interaction tensor $\mc T_\mathsf i$, are elementary symmetries of the tensor network~\cite{PhysRevB.97.125102}. 
This implies that only the parity of the number of $X$ operators pushed onto a qubit vertex, and the parity of the number of $Z$ operators pushed onto an interaction vertex, need to be accounted for. 
Moreover, any product of $X$ operators on qubit vertices in the image of the bipartite adjacency matrix $\varsigma_X$ is a virtual symmetry. 
In fact these operators can be moved to the physical legs of the tensor network, where they form operators of the stabilizer group. 

Starting from the tensor network defined above, a spanning set of ground state tensor networks can be generated by moving $\overline{X}$ \emph{logical operators} given by representatives of equivalence classes ${\rm ker} \, \varsigma_Z^\dagger / {\rm im}\, \varsigma_X$ to the virtual level, where $\varsigma_Z : \mathbb Z_2[\mc I^Z] \to \mathbb Z_2[\mc S]$ is defined similarly to $\varsigma_X$. These tensor networks having topological order, they are locally indistinguishable, i.e. none of the logical operators can be supported within a ball. Finally, note that no virtual $Z$ operator on an interaction vertex is a virtual symmetry. Indeed, all such operators generate orthogonal tensor network states, as they correspond to flipping the eigenvalue of an operator $\mathbb B_\mathsf i (X)$ from $+1$ to $-1$.

\subsection{Topological isometry condition}

\noindent
We shall now discuss a characteristic property of the topological tensor network states defined in the previous part, namely the \emph{isometry} condition. We begin with a minimal example. Every tensors $\mc T_\mathsf s$ associated with a site $\mathsf s$ satisfies
\begin{align}
    \mc T_\mathsf s^\dagger \mc T_\mathsf s = 
    \delta \Big(\sum_{\mathsf i \in N_\mathsf s}a_{\mathsf s, \mathsf i} \Big) \bigotimes_{\mathsf i\in N_\mathsf s} X_{\mathsf s,\mathsf i}^{a_{\mathsf s, \mathsf i}}
    \, ,
\end{align}
where as before repeated indices are contracted. It follows that the tensor $\mc T_\mathsf s$ are \emph{isometric}---and a fortiori \emph{injective}---on the support subspace of the unnormalized projector $\mc T^\dagger_\mathsf s \mc T_\mathsf s$. Given a topological stabilizer tensor network as previously defined, we always have the possibility of concatenating to $\mc T_\mathsf s$ adjacent $\delta$ tensors, providing to the state a more conventional PEPS structure, where non-trivial tensors only sit at the sites $\mathsf s$. This has the simple effect of introducing further redundant virtual legs that are enforced to be in a common state, modifying in a trivial way the injectivity condition.

Although the symmetry and injectivity subspace of a single $\mc T_\mathsf s$ tensor is very simple, for larger regions, the connectivity of the tensor network can lead to very non-trivial constraints capable of describing any topological stabilizer code. 
Given a topological tensor network, let us consider a large region $\mc R$, and its complement $\mc R^{\rm c}$, containing many qubit sites $\mathsf s$. We partition the set of interaction vertices $\mathsf i$ into those in $\mc I^X_\mc R$ or $\mc I^X_{\mc R^{\rm c}}$ that satisfy $N_\mathsf i \subseteq \mc R$ or $N_\mathsf i \subseteq \mc R^{\rm c}$, respectively, and those on the boundary in $\mc I^X_{\partial \mc R}$ that satisfy $N_\mathsf i \not\subseteq \mc R$ and  $N_\mathsf i \not\subseteq \mc R^{\rm c}$. By definition, the interaction vertices in $\mc I^X_{\partial \mc R}$ are connected both with qubits in $\mc R$ and in the complement $\mc R^{\rm c}$.  
For these boundary interaction vertices, we define the interaction tensor restricted to $\mc R$ as 
\begin{align}
    \mc T_\mathsf i {\sss |}_{\mc R} =
    \Big( \prod_{\mathsf s  \in N_\mathsf i \cap \mc R}\delta(a_{ \mathsf i } + a_{\mathsf s,\mathsf i}) \Big)
	\, | \{a_{\mathsf s , \mathsf i}\}_{\mathsf s \in N_\mathsf i \cap \mc R}\ra \la a_\mathsf i | \, .
\end{align}
so that the tensor network associated with $\mc R$ reads
\begin{align}
    \mc T_\mc R= 
    {\rm tr}\Big(\bigotimes_{\mathsf  s \in \mc R} \mc T_\mathsf s
    \bigotimes_{\mathsf i \in \mc I^X_\mc R} \mc T_\mathsf i
    \bigotimes_{\mathsf i \in \mc I^X_{\partial \mc R}} \mc T_\mathsf i {\sss |}_{\mc R}\Big)
    \, .
\end{align}
In other words, we construct the tensor network associated with the whole interaction graph proceeding as previously, and then split all the $\mc T_\mathsf i$ tensors with $\mathsf i \in \mc I^X_{\partial \mc R}$ between two tensors $\mc T_\mathsf i {\sss |}_\mc R$ and $\mc T_\mathsf i {\sss |}_{\mc R^{\rm c}}$, e.g.
\begin{equation*}
    \includegraphics[scale=1, valign=c]{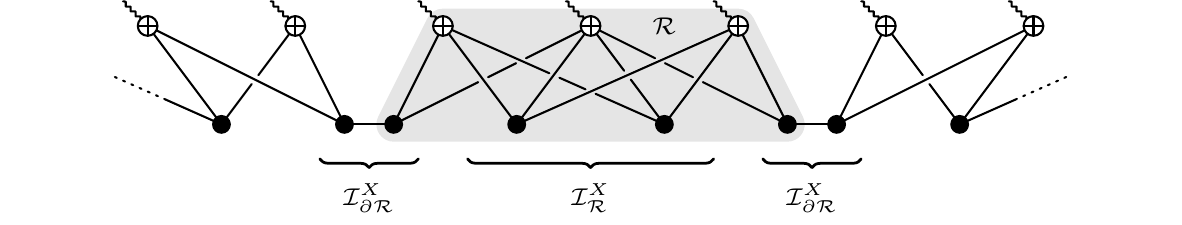} \hspace{-3em} ,
\end{equation*}
such that contracting all the tensors in the gray area yields $\mc T_\mc R$.

The tensor $\mc T_\mc R$ is a map from the boundary Hilbert space, which has one qubit per interaction vertex $\mathsf i \in \mc I^X_{\partial \mc R}$ on the boundary, to the bulk Hilbert space, which consists of the bulk qubits $\{\mathsf s \in \mc R\}$. 
The virtual symmetries that leave the tensor $\mc T_\mc R$ invariant are given by elements in $\ker (\varsigma_X {\sss |}_\mc R)$, where $\varsigma_X {\sss |}_\mc R$ is the bipartite adjacency matrix $\varsigma_X$ with the input restricted to interaction vertices $\{\mathsf i \in \mc I^X_\mc R\cup \mc I^X_{\partial \mc R} \}$ and the output restricted to qubits $\{\mathsf s \in \mc R \}$. 
We further partition the virtual symmetries into the set of \textit{local relations} given by $\ker (\varsigma_X {\sss |}_{\mc R \cup \partial \mc R})$, where $\varsigma_X {\sss |}_{\mc R \cup \partial \mc R}$ denotes $\varsigma_X $ with input restricted to $\mc I^X_\mc R\cup \mc I^X_{\partial \mc R}$ and no output restrictions, 
and the boundary symmetries 
\begin{align}
    \mathcal{B}_\mc R = \frac{\ker (\varsigma_X {\sss |}_\mc R)}{\ker (\varsigma_X {\sss |}_{\mc R \cup \partial \mc R})}
    \, ,
\end{align}
which are defined up to local relations. 
The elements of $\mathcal{B}_\mc R$ are boundary symmetry operators that are labelled by vectors $\mathfrak b \in \mathbb{Z}_2[\mc I^X_{\partial \mc R}]$ 
\begin{align}
    X(\mathfrak b) = \bigotimes_{\mathsf i\in \mc I^X_{\partial \mc R}} X_\mathsf i^{\mathfrak b_\mathsf i} 
    \, ,
\end{align}
that satisfy $\mc T_\mc R \mathfrak b = \mc T_\mc R$. 
The tensor $\mc T_\mc R$ is then isometric on the symmetric subspace of $\mathcal{B}_\mc R$
\begin{align}
    \mc T_\mc R^\dagger \mc T_\mc R = \sum_{\mathfrak b \in \mathcal{B}_\mc R} X(\mathfrak b)
    \, .
\end{align}
An example of the procedure described above is as follows: Given the 2-form representation of the 3d toric code discussed in the main text, we consider the region $\mc R$ that includes the six qubits---to which are associated $\delta^\oplus$ tensors---surrounding a vertex of the underlying cubic cellulation. The twelve $\delta$ tensors associated with the surrounding plaquettes are associated with interaction vertices on the boundary of $\mc R$. Let us split these four-valent $\delta$ tensors into pairs of three-valent ones, such that one tensor in every pair is connected to two $\delta^\oplus$ tensors within $\mc R$. The $\delta$ tensors satisfying this criterion correspond to the boundary interaction tensors restricted to $\mc R$ introduced above. Contracting these twelve boundary tensors with the six $\delta^\oplus$ tensors in $\mc R$ results in a twelve-valent isometric tensor, whose properties are discussed in detail in \cite{delcamp2020tensor}.

\subsection{Topological excitations}

\noindent
States containing topological charge excitations can be built from the ground state tensor network by including virtual operators. 
These come in two types. The first one arises when including a truncated virtual symmetry $b_{\text{trunc}}$, which creates some excitations along the boundary where the symmetry has been truncated. These excitations may appear along the whole boundary, such as a loop excitation in the 3d toric code, or along a collection of points within the boundary, such as in a fracton model. Any truncated virtual symmetry can be directly lifted to the physical level and hence reflects the physical operator that creates the same excitation pattern. We remark that the distinct ground state tensor networks obtained by moving the logical $\overline{X}$ operators to the virtual level can be interpreted as nucleating such an excitation pattern over a non-contractible cycle. 
The second type of excitations is obtained by inserting $Z$ operators on the virtual level. Only products of $Z$ operators on a set of interaction vertices in the image of the adjoint bipartite adjacency matrix $\im \varsigma_X^\dagger$ can be lifted to the physical level. 
Products of $Z$ operators orthogonal to this image, i.e. in $(\im  \varsigma_X^\dagger)^\perp$, correspond to non-trivial topological charges. 
On a region $\mc R$, the topological super-selection sectors of these charges are given by  $\{\mathsf i \in \mc I^X_\mc R \cup \mc I^X_{\partial\mc R} \}/\im  \varsigma_X^\dagger$, which are  one-to-one with the irreducible representations of the boundary symmetries $\mathcal{B}_\mc R$. 
These operators can be thought of as measuring the flux through the boundary due to the charges in the bulk via a generalized Gau{\ss}'s law. 
Due to the form of $\mathcal{B}_\mc R$, any allowed charge configuration (see below) in $\mc R$ can be neutralized via a charge configuration on the boundary. 

There are further constraints on the \textit{allowed} charge configurations that do not force the tensor network to vanish. 
These are derived from \textit{constraints}, or \textit{materialized symmetries}~\cite{KITAEV20032}, captured by $\ker \varsigma_X$, whose elements leave not just the ground state, but the exact form of the tensor network itself invariant. 
Allowed charge configurations must be even under all relations, otherwise they result in a vanishing state. 
The relations can be decomposed into local relations $\ker (\varsigma_X {\sss |}_{\mc R \cup \partial \mc R})$ 
contained within topologically trivial regions $\mc R$ and global relations $\ker \varsigma_X / \{ \text{local relations} \}$ that appear once the boundary conditions have been fixed, possibly forming a non-trivial topology. For example, the 3d toric code has \emph{local} constraints forcing the string-like excitations to appear in closed loops, it also has a \emph{global} constraint for closed boundary conditions that forces the number of point-like charges to be even. Fracton models in 3d, on the other hand, possess only global constraints.

\subsection{Tensor perturbations}

\noindent
Perturbations to the tensor network are generated by $X$ and $Z$ operators on the $\mc T_\mathsf s$ and $\mc T_\mathsf i$ tensors, respectively, as mentioned above. 
Given the properties of the $\delta^\oplus$ tensors, a uniform $X$ perturbation is identical to a uniform $X$ perturbation on the physical level, and hence any tensor network in a gapped topological phase is stable to these perturbations. 
However, we explained that $Z$ perturbations correspond to non-trivial topological charges, and similarly for uniform $Y$ perturbations, as noted in the main text.
Therefore, a uniform $Z$ tensor perturbation is not equivalent to a uniform local perturbation on the physical spins, and so there is no guarantee of stability against such perturbations. 
In fact, such perturbations induce fluctuation of the topological charges represented by virtual $Z$ operators, which can lead to \emph{instability}. For example, all 2d MPO-injective tensor networks are unstable to generic uniform tensor perturbations as they induce anyon condensation phase transitions. 
As argued in the main text, the robustness of a tensor network to perturbations depends crucially on the type of charges fluctuated as well as the corresponding constraints, or materialized symmetries, they satisfy. We shall now demonstrate this proposal for a general CSS tensor network state.

\begin{figure}[t]
    \centering
	\includegraphics[scale=1]{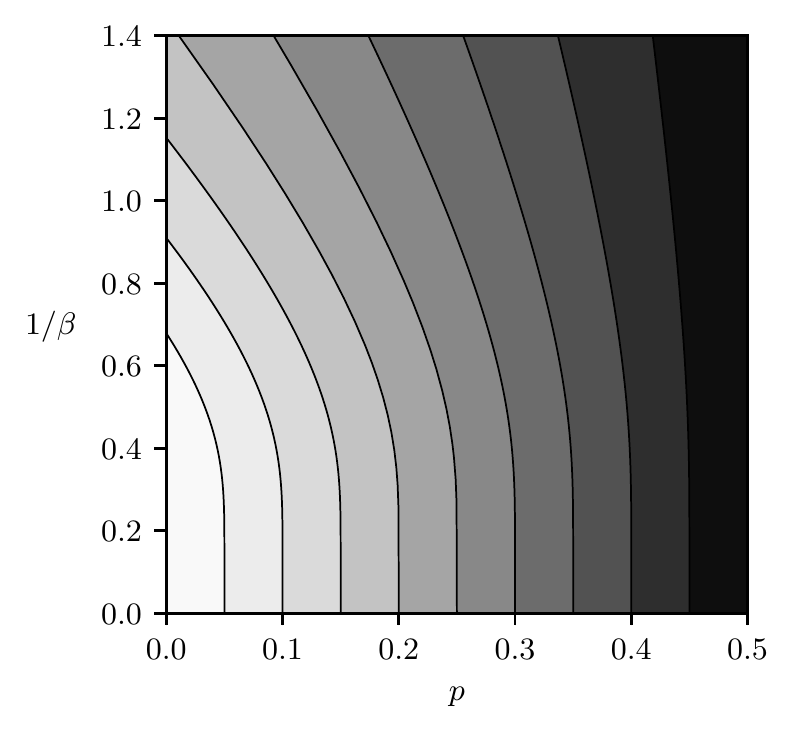}
    \caption{ A contour plot showing lines in the temperature $\nicefrac{1}{\beta}$, disorder probability $p$, plane that satisfy  Eq.~\eqref{eq:munubp} for fixed $\mu$ and $\nu$. Varying the normalization $c$ moves the model along such a line.}
\label{fig:Contour}
\end{figure}

We focus on a uniform $Z$ perturbation to all $\mc T_\mathsf i$ tensors, writing the unnormalized perturbed tensor as 
\begin{align}
    \label{eq:pertubGen}
    \mc T_\mathsf i \mapsto \mu \mc T_\mathsf i + \nu \widetilde{\mc T}_\mathsf i
    \, , 
\end{align}
where  $\widetilde{ \mc T}_\mathsf i$ denotes $\mc T_\mathsf i$ multiplied by a single $Z$ operator, and $|\nu| \ll |\mu|$ for a small perturbation. 
Denoting the perturbed tensor network state by $\ket{\mu,\nu}$, we can use the properties of the $\delta$ and $\delta^\oplus$ tensors outlined in the main text to show that its norm depends only on the real parameters $|\mu|^2$ and $|\nu^2|$:
\begin{align}
    \la \mu,\nu | \mu , \nu\ra \propto \bigotimes_{\mathsf s \in \mc S} \big( \la 0 |\mc T_\mathsf{s}\big) \; \bigotimes_{\mathsf i \in \mc I^X}
    \big(|\mu|^2 \mc T_\mathsf{i} + |\nu|^2 \widetilde{\mc T}_\mathsf{i} \big)
    \, ,
\end{align}
up to an overall normalization. 
Shifting the overall normalization of the state by $\nicefrac{1}{\sqrt{c}}$ for each interaction vertex, we obtain $\ket{\nicefrac{\mu}{\sqrt{c}},\nicefrac{\nu}{\sqrt{c}}}$, whose norm can be mapped to the partitition function of a \emph{generalized} classical Ising model with \emph{disorder}. 
To this end, we take the overall normalization factor $c$ to satisfy the equations 
\begin{align}
    \frac{|\mu|^2}{c} = (1-p) e^{\beta} + p e^{-\beta} \q {\rm and} \q
    \frac{|\nu|^2}{c} = (1-p) e^{-\beta} + p e^{\beta} \, .
\end{align}
Shifting $c$ moves the norm, as a function of $|\mu|^2$ and $|\nu|^2$, with fixed $\mu$ and $\nu$ along the line in the $(\beta,p)$ plane that satisfies 
\begin{align}
    \frac{|\mu|^2-|\nu|^2}{|\mu|^2+|\nu|^2} &= (1-2p) \tanh(\beta) \, .
    \label{eq:munubp}
\end{align}
Such lines are parameterized by $c$ as follows (see Fig.~\ref{fig:Contour}):
\begin{align}
    \frac{|\mu|^2+|\nu|^2}{c} =e^\beta+e^{-\beta} \q , \q
    \frac{|\mu|^2|\nu|^2-c^2}{(|\mu|^2+|\nu|^2)^2-4 c^2} = p(1-p)
     \, .
\end{align}
This allows us to pick a normalization factor $c$ corresponding to the most convenient mix of temperature and disorder for our purposes, in particular we restrict to zero disorder below. 

Performing a Fourier transform via the introduction of Hadamard matrices, we obtain
\begin{align}
    \nn
    \Big\la \frac{\mu}{\sqrt c}, \frac{\nu}{\sqrt c} \Big| \frac{\mu}{\sqrt c}, \frac{\nu}{\sqrt c} \Big\ra
    &\propto \bigotimes_{\mathsf s \in \mc S} 
    \big( \la 0 | \mc T_\mathsf s H^{\otimes |N_\mathsf s|}\big) \;
    \bigotimes_{\mathsf i \in \mc I^X} \Big( \big( (1-p) e^{\beta} + p e^{-\beta} \big) H^{\otimes |N_\mathsf i|} \mc T_\mathsf i + \big( (1-p) e^{-\beta} + p e^{\beta} \big) H^{\otimes |N_\mathsf i|} \widetilde{\mc T}_\mathsf i \Big)
    \\
    \nn
     &\propto \sum_{\{\sigma = \pm 1\}}\sum_{\{\eta = \pm 1\}} \prod_{\mathsf i \in \mc I^X} p^{\frac{1-\eta_\mathsf i}{2}} (1-p)^{\frac{1+\eta_\mathsf i}{2}} \bigg( e^{\beta} \delta \Big(\eta_\mathsf i \prod_{\mathsf s \in N_\mathsf i} \sigma_\mathsf s =1 \Big) + e^{- \beta} \delta \Big(\eta_\mathsf i \prod_{\mathsf s\in N_\mathsf i} \sigma_\mathsf s  =-1 \Big) \bigg)
    \\
     &\propto \sum_{\eta} \text{Prob}(\eta) \, \mc Z[\beta,\eta]
     =
     \mc Z[\beta,p] \, ,
\end{align}
where $\eta \equiv \{\eta_\mathsf i \}_{\mathsf i \in \mc I^X}$ is a set of \emph{independent} and \emph{identically distributed} random variables with probability ${\text{Prob}(\eta_\mathsf i = -1)=p}$ and ${\text{Prob}(\eta_\mathsf i=+1)=1-p}$, and the partition function with fixed disorder is given by
\begin{align}
    \mc Z[\beta,\eta] &=  \sum_{\{\sigma = \pm 1\}} \exp \Big( \beta \sum_{\mathsf i \in \mc I^X} \eta_\mathsf i \prod_{\mathsf s \in N_\mathsf i} \sigma_\mathsf s \Big) \, .
\end{align}
In addition to a change of basis on the internal indices of the tensor network, we used in the above that $\la 0 | \mc T_\mathsf s H^{\otimes |N_\mathsf s|}$ and $H^{\otimes |N_\mathsf i|} \mc T_\mathsf i$ are proportional to a $\delta$  and $\delta^\oplus$ tensor, respectively.
Moreover, the classical spin variables are given by $X$-basis states $\sigma_\mathsf s = \pm 1$. 
The partition function corresponds to a generalized classical Ising model with disorder strength $p$, at inverse temperature $\beta$ \cite{doi:10.1063/1.1499754}. 
In particular, the two dimensional phase diagram of this disordered model, as a function of inverse temperature and disorder strength, contains the \textit{Nishimori line} along which $e^{-2\beta} = \frac{p}{1-p}$ \cite{10.1143/PTP.66.1169}.

Futhermore, we remark that 
\begin{align}
    \mc Z[\beta,\eta+\zeta] = \mc Z[\beta,\eta] \, ,
\end{align}
for a disorder configuration $\zeta$ corresponding to a neutral cluster of charges. 
A cluster of charges within a ball-like region $\mc R$ is \emph{neutral} if it is symmetric under all virtual symmetries on the boundary of the region $\partial \mc R$. 
A neutral cluster can be created by a local operator within $\mc R$ on the physical level. For the CSS tensor networks we consider, this operator is a product of Pauli-$Z$ matrices, without loss of generality. 
Since CSS tensor networks containing different excitations are orthogonal, the same neutral cluster must appear in the bra and ket layers of the norm. Both can be moved to operators on the physical level, which multiply to give a stabilizer that is a symmetry of the state. 
Hence, a disorder configuration $\zeta$ corresponding to a neutral cluster is a symmetry of the partition function in the sense above. 
This implies that the value of the partition function depends only on equivalence classes of disorder  $[\eta]=\{ \eta + \zeta \}_\zeta$ and so we can write $Z[\beta,[\eta]]$. 
Therefore, one has 
\begin{align}
    \mc Z[\beta,p] &= \sum_{[\eta]} \text{Prob}([\eta])\, \mc Z[\beta,[\eta]] 
    \, ,
\end{align}
where $\text{Prob}([\eta])=\sum_{\eta' \in [\eta]} \text{Prob}(\eta')$.
We further notice that $\mc Z[\infty,\eta]= 0$ if $\eta$ violates any local materialized symmetries, which are independent of the boundary conditions, and also if $\eta$ violates some global materialized symmetry once boundary conditions have been fixed. In other words,  $\mc Z[\infty,\eta]=0$ whenever $[\eta]\neq 1$. 

Importantly, Pauli-$X$ operators on the quantum spins map directly to classical spin operators 
\begin{align}
    \frac{\left\langle\frac{\mu}{\sqrt c}, \frac{\nu}{\sqrt c}  \right| \prod_{\mathsf s \in \mc C} X_\mathsf s \left| \frac{\mu}{\sqrt c}, \frac{\nu}{\sqrt c} \right\rangle}
    {\left\langle\frac{\mu}{\sqrt c}, \frac{\nu}{\sqrt c} \middle| \frac{\mu}{\sqrt c}, \frac{\nu}{\sqrt c}\right\rangle}
    &= 
    \bigg\la
    \Big\la \prod_{\mathsf s\in \mc C} \sigma_\mathsf s \Big\ra_{\! \beta}
    \bigg\ra_{\! p}
    \, ,
\end{align}
where $\mc C$ denotes here a collection of spins, whereas Pauli-$Z$ operators map to classical spin flip operators
\begin{align}
    Z(a) : \sigma_\mathsf s \mapsto (-1)^{\mathfrak a_\mathsf s} \sigma_\mathsf s
    \, ,
\end{align}
where $\mathfrak a\in\mathbb{Z}_2[\mc S]$. The zero temperature and zero disorder limit, i.e. $\beta\rightarrow \infty$ and $p\rightarrow 0$, of the partititon function corresponds to the norm of the unperturbed tensor network. This simply results in an equal weighted sum of allowed spin configurations
\begin{align}
    \mc Z[\infty,0]
    & \propto \sum_{\{\sigma_\mathsf s\}} \prod_{\mathsf i \in \mc I^X} \delta\Big( \prod_{\mathsf s \in N_\mathsf i} \sigma_\mathsf s = 1 \Big)
    \, ,
\end{align}
up to an overall normalization. 
In this limit the virtual symmetries (equivalently the $X$ stabilizers) map to symmetries of the classical model
\begin{align}
    \label{eq:XStabilizerToClassical}
    \bigg\la
    \Big\la \prod_{\mathsf s\in \mathfrak b} \sigma_\mathsf s \Big\ra_{\! \beta}
    \bigg\ra_{\! p} = 1 \, ,
\end{align}
for $\mathfrak b \in \im \varsigma_X$. At non-zero temperature and disorder strength, these symmetries are broken by the tensor perturbations as they carry non-trivial topological charges. 
On the other hand, the Pauli-$Z$ stabilizers (and logical operators) ${Z(\mathfrak a),~\mathfrak a\in\ker \varsigma_X^\dagger,}$ map to spin flip symmetries of the classical model 
\begin{align}
    \exp \Big( \beta \sum_{\mathsf i \in \mc I^X} \prod_{\mathsf s \in N_\mathsf i} (-1)^{\mathfrak a_\mathsf s} \sigma_s \Big) =  \exp \Big( \beta \sum_{\mathsf i \in \mc I^X} \prod_{\mathsf s \in N_\mathsf i} \sigma_\mathsf s \Big)
    \, ,
\end{align}
which persist to arbitrary $p$ and $\beta$. 
Relations on the $Z$ stabilizers lead to relations on the classical spin flip symmetries, i.e. nontrivial products of spin flips that act trivially on the classical spins. 
Hence the  classical partition function can be viewed as a generalized disordered Ising model with spin flip symmetries given by $\ker \varsigma_X^\dagger$. 
The non-conserved operators from Eq.~\eqref{eq:XStabilizerToClassical} in fact become order parameters for a phase transition of the classical model. 

Due to the concavity of $\log$, the order parameter with finite disorder only provides an upper bound on the value of the true quantum order parameter
\begin{align}
    - \log \Big\la \prod_{\mathsf s \in \mc C} X_\mathsf s \Big\ra_{ \! {\frac{\mu}{\sqrt c}, \frac{\nu}{\sqrt c}}}
    &\leq 
    \bigg\la \!\!
    - \log
    \Big\la \prod_{\mathsf s\in \mc C} \sigma_\mathsf s \Big\ra_{\! \beta}
    \bigg \rangle_{\! p}
    \, ,
\end{align}
for $p>0$. Hence the phase boundaries of the quantum and classical models need not match at finite disorder strength, rather the phase transition line for the disordered classical model falls within the quantum phase that originates at zero temperature. 
For this reason we focus on the $p=0$ case below and in the main text.

With no disorder $p=0$ we see that the quantum phase diagram of the perturbed tensor network representation reduces to the finite temperature phase diagram of a generalized classical Ising model. 
Hence if the classical model has a zero temperature phase transition, the tensor network is not stable to perturbations. On the other hand, if the classical model has no zero temperature phase transition, the tensor network is stable to sufficiently small perturbations, provided the tensor network state has a gapped parent Hamiltonian with topological order to also guarantee stability to perturbations that can be locally lifted to the physical level. 
Tensor networks that lead to classical models with a zero temperature phase transition, such as 3d fracton topological orders, or a toric code tensor network with 1-form virtual symmetry, are unstable. 
While tensor networks that lead to classical models with non-zero temperature phase transition, such as toric code tensor networks with 2-form (and higher) virtual symmetries, are stable.

The stability criteria obtained above extends to general local perturbations, as they can be decomposed into virtual components given by $X$ and $Z$ operators, respectively. We have already established that the tensor network state is stable to $X$ perturbations due to the inherent topological robustness of the phase. If the tensor network state is additionally stable to $Z$ perturbations, as per the above procedure, we expect stability to general uniform perturbations.
For instance, as explained in the main text in the context of the toric code, a uniform $Y$ perturbation on the virtual level leads to the same classical partition function as a uniform virtual $Z$ perturbation. Therefore, (in)stability to $Z$ perturbations extends to $Y$ perturbations.
More generally, a perturbation of the form 
\begin{align}
    \mc T_\mathsf{i} \to \mu \mc T_\mathsf{i} + \nu  \widetilde{\mc T}_\mathsf{i} + \xi X \mc T_\mathsf{i} + \zeta X \widetilde{\mc T}_\mathsf{i}
    \, , 
\end{align}
can be written as a physical perturbation applied to the perturbed tensor network that results from the modification considered in Eq.~\eqref{eq:pertubGen}. In this case, the physical perturbation applies either a $(\mu\openone + \xi X)$ or $( \openone + \zeta /\nu X)$ operator controlled by the absence or presence of a string excitation segment at that point, respectively. This argument requires $|\zeta| \ll |\xi|,|\nu| \ll |\mu|$, but we suspect $|\zeta|,|\xi|,|\nu| \ll |\mu|$ suffices as in the case of a uniform $Y$ perturbation. 
We remark that the perturbation considered above involves a generating set for the full algebra of virtual operators. We expect that stability to such uniform perturbations is indicative of stability to arbitrary local perturbations, i.e. those with exponentially decaying correlations in space. 

\medskip \noindent
In the discussion of the main text, we established a conjectural correspondence between the stability of tensor network representations and the topology of the corresponding virtual charges. We expect this correspondence to hold for more general topological quantum liquids, i.e. those that depend only on the topology of the manifold. Akin to the 1- and 2-form representations of the toric code, we should be able to define two tensor network representations that satisfy virtual symmetries with respect to membrane-like and string-like projected entangled-pair operators (PEPO), respectively. The latter representation whose virtual charges are mapped to the loop-like physical excitations is expected to be stable in general. However, the proof of the stability in this more general scenario is more subtle. In particular, the identification between bra and ket layers that occurs when considering the norm of the perturbed tensor network of a CSS model will not hold. Instead, in the simple case where distinct excited states built on top of the tensor network are taken to be orthogonal, the product of a given symmetry operator in the bra and ket is again a symmetry of the perturbed tensor network. This effectively leaves only a single independent copy of the virtual symmetry, similar to the stabilizer examples, which is then explicitly broken by tensor perturbations.

In higher dimension there exist topological quantum liquid tensor network states with higher form virtual symmetries from 1- up to ($d-$1)-form. We anticipate that all those with 2-form or higher virtual symmetry should be stable. For instance, the 4d 2-form $\mathbb Z_2$ gauge theory (sometimes referred to as the 4d toric code, see e.g.~\cite{doi:10.1063/1.1499754}) admits two tensor network representations with 2-form virtual symmetry that are both stable to perturbations.
We further remark that these higher form symmetric topological quantum liquid tensor networks should fall under the generalized tensor network injectivity formalism based on state-sum topological quantum field theories, but it is possible to develop the theory of specific higher form symmetry cases beyond this general formalism. Finally, it would also be interesting to study the virtual symmetries of more general non-liquid fracton topological orders, including those with non-abelian particles.

\subsection{A non-topological example: the 2d quantum Ising model}

\noindent
We presented above a general recipe to assess the stability of a given topological tensor network state under arbitrary perturbations. In order to illustrate the fact that our stability criterion is specific to tensor network states that have gapped parent Hamiltonians with topological order, we shall now consider a non-topological example, namely the 2d quantum Ising model. This model is defined in terms of spin variables on the vertices of a square lattice that are governed by the Hamiltonian 
\begin{align}
    \mathbb H_{\rm Ising} = -\sum_{\la \mathsf u \, \mathsf v \ra} X_\mathsf u X_\mathsf v
    \, ,
\end{align}
where $\la \mathsf u \, \mathsf v\ra$ denotes nearest neighbouring vertices.
Note that we  have chosen the basis that is consistent with the presentation above. 
The tensor network for the ground state is obtained by contracting local tensors associated with every vertex $\mathsf v$ and every pair $\la \mathsf u \, \mathsf v \ra$ such that the unit cell reads
\begin{equation}
    \includegraphics[scale=1, valign=c]{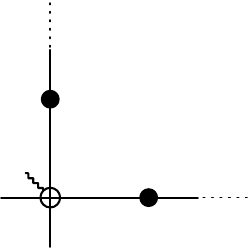} \, .
\end{equation}
This tensor network actually has a 2-form virtual symmetry, which in this case corresponds to the fact that any pair of $X$ operators on virtual indices leaves the state invariant. Creating a pair of virtual $X$ operators, running them over a closed loop, and annihilating them, generates a 1-form relation. 
Additionally, $X$ perturbations on the virtual level map directly to physical perturbations, while $Z$ perturbations must satisfy the 1-form relation by  appearing in closed loops on the dual lattice, otherwise the perturbed tensor network evaluates to zero. 
Unsurprisingly, the norm of the tensor network with $Z$ perturbations, i.e. $\mc T_{\langle \mathsf u \mathsf v\rangle} \mapsto \mc T_{\langle \mathsf u \mathsf v\rangle} + \widetilde{\mc T}_{\langle \mathsf u \mathsf v\rangle}$, reproduces the disordered 2d classical Ising model 
\begin{align}
    \braket{\mu, \nu | \mu, \nu}  \propto 
    \sum_{\eta} \prod_\mathsf e p^\frac{1-\eta_\mathsf e}{2}(1-p)^\frac{1+\eta_\mathsf e}{2} \sum_{\{\sigma = \pm 1\}} {\rm exp} \Big( \beta \sum_{\langle \mathsf u \,  \mathsf v \rangle} \eta_{\langle \mathsf u \, \mathsf v \rangle}  \sigma_\mathsf u \sigma_ \mathsf v \Big)
    \, , 
\end{align}
up to an overall normalization constant, where $\mu$ and $\nu$ are related to $\beta$ and $p$ via Eq.~\eqref{eq:munubp}. 
For $|\nu|\ll |\mu|$, this corresponds to the ordered phase of the classical Ising model i.e.
    \begin{align}
    \bra{\mu, \nu} X_\mathsf v X_{\mathsf v+\hat x} \ket{\mu, \nu}  = 
    \big\la \la \sigma_\mathsf v \sigma_{\mathsf v+ \hat x} \ra_\beta \big\ra_p
    \xrightarrow[| \hat x | \to \infty]{} 
    \text{const.} \, .
\end{align}
However, this does not mean that this tensor network is stable to perturbations. Indeed, virtual $X$ perturbations that map directly to physical perturbations can cause a phase transition at infinitesimal strength since the Ising model does not have topological order and the $X$ perturbation explicitly breaks the global $\mathbb Z_2$ symmetry of the model. This is in sharp contrast to the tensor networks with topological order considered in the main text.

\section{Finite temperature quantum phase mapping \label{sec:app_ft}}

\noindent
In this section, we consider the quantum partition function of a general CSS code of the form
\begin{equation}
    \label{eq:CSSHam}
    \mathbb H[\cubu] = \lambda_\mathbb A \mathbb H_\mathbb A(Z) + \lambda_\mathbb B \mathbb H_\mathbb B(X) \, .
\end{equation}
We begin by showing that the quantum partition function of $\mathbb H[\cubu]$ factorizes into the product of two terms associated with the quantum partition functions of $\mathbb H[\cubu]$ for $\lambda_\mathbb A=0$ and $\lambda_\mathbb B=0$, respectively. We then demonstrate that the norm of a perturbed tensor network state under local (uniform) perturbations can always be mapped to that of the quantum partition function of a CSS Hamiltonian of the form \eqref{eq:CSSHam} for $\lambda_\mathbb A=0$ or $\lambda_\mathbb B=0$.

Let us consider the quantum partition function $\mc Z[\beta] = {\rm tr}(e^{-\beta \mathbb H[\cubu]})$ of the CSS Hamiltonian. 
Denoting by ${|\{ a =0,1 \} \ra}$ and ${| \{ \sigma = \pm \} \ra}$ microscopic states, where $\{a\}$ and $\{\sigma\}$ are configurations for all the spins of the lattice expressed in the Pauli-$Z$ and -$X$ bases, respectively, we have
\begin{align}
    e^{-\beta \mathbb H[\cubu]} 
    = \sum_{\substack{\{a=0,1\} \\ \{\sigma=\pm\}}}
    |\{a\} \ra
    e^{- \beta \lambda_\mathbb A \la \{a\}|\mathbb H_\mathbb A(Z)|\{a\} \ra }
    \la \{a\} | \{\sigma\} \ra
    e^{-\beta \lambda_\mathbb B \la \{\sigma\} |\mathbb H_\mathbb B(X) | \{\sigma\} \ra}   \la \{\sigma\} | \, ,
\end{align}
where we used the fact that $\mathbb H_\mathbb A(Z)$ and $\mathbb H_\mathbb B(X)$ are diagonal in the Pauli-$Z$ and -$X$ bases, respectively.
Since $\la \{a\}|\{\sigma\} \ra^2$ is independent of both $\{a\}$ and $\{\sigma\}$, we obtain that the quantum partition function factorizes as
\begin{align}
    \mathcal Z[\beta] = 
    \sum_{\substack{\{a=0,1\} \\ \{\sigma=\pm\}}} 
    e^{- \beta \lambda_\mathbb A \la \{a\}|\mathbb H_\mathbb A(Z)|\{a\} \ra }
    \la \{a\} | \{\sigma\} \ra^2
    e^{-\beta \lambda_\mathbb B \la \{\sigma\} |\mathbb H_\mathbb B(X) | \{\sigma\} \ra}  
    \ \propto \, \mathcal Z_{\mathbb A}[\beta] \cdot \mathcal Z_\mathbb B[\beta] \, ,
\end{align}
such that $\mc Z_\mathbb A[\beta] = {\rm tr}(e^{-\beta \lambda_\mathbb  A \mathbb H_\mathbb A(Z)})$, $\mc Z_\mathbb B[\beta] = {\rm tr}(e^{-\beta \lambda_\mathbb B \mathbb H_\mathbb B(X)})$ and the proportionality constant is independent of $\beta$.

Given a CSS Hamiltonian \eqref{eq:CSSHam} with 
\begin{equation}
    \mathbb H_\mathbb B(X) = -\sum_{\mathsf i \in \mc I^X} \mathbb B_\mathsf i(X)
    \q\q {\rm with} \q \mathbb B_\mathsf i (X)= \prod_{\mathsf s \in N_\mathsf i} X_\mathsf s \, ,
\end{equation}
we constructed in App.~\ref{sec:app_topoTNCSS} a tensor network representation of the ground state 
\begin{equation}
    \Big(\prod_{\mathsf i \in \mc I^X} \frac{1}{2} \big({\rm id}+\mathbb B_\mathsf i (X) \big)\Big)
    | 0\ra^{\otimes |\mc S|}
\end{equation}
such that the norm of the perturbed tensor network can be mapped to the following classical partition function:
\begin{equation}
    \sum_{\{\sigma = \pm 1 \}} \! \prod_{\mathsf i \in \mc I^X} {\rm exp}\Big( \beta \prod_{\mathsf s \in N_\mathsf i } \sigma_\mathsf s\Big) \, ,
\end{equation}
namely a generalized classical Ising model without disorder ($p \to 0$).
It turns out that the classical partition above corresponds to the quantum partition function of $\mathbb H[\cubu]$ for $\lambda_\mathbb A = 0$ and $\lambda_\mathbb B= 1$. Indeed, since $[\mathbb B_\mathsf i, \mathbb B_{\mathsf i'}]=0$ for every $\mathsf i, 
\mathsf i'$, we have
\begin{equation}
    \mc Z_\mathbb B[\beta] = 
    \sum_{\{\sigma = \pm\}}
    e^{-\beta \la \{\sigma\}|\mathbb H_\mathbb B(X)|\{\sigma\}\ra}
    =
    \sum_{\{\sigma=\pm\}} \prod_{\mathsf i \in \mc I^X}
    e^{\beta \la \{\sigma\}|\mathbb B_\mathsf i(X) | \{\sigma\} \ra} \, .
\end{equation}
Writing $\la \{\sigma\}|\mathbb B_\mathsf i(X)|\{\sigma\}\ra$ explicitly, it follows that
\begin{align}
    \mc Z_\mathbb B[\beta] 
    = \sum_{\{\sigma=\pm\}}
    \prod_{\mathsf i \in \mc I^X}
    {\rm exp}\Big( \beta \big\la \{\sigma_\mathsf s\}_{\mathsf s \in N_\mathsf i} \big| \prod_{\mathsf s \in N_\mathsf i} X_\mathsf s \big| \{\sigma_\mathsf s\}_{\mathsf s \in N_\mathsf i} \big\ra \Big)
    =  \sum_{\{\sigma = \pm 1 \}} \!
    \prod_{\mathsf i \in \mc I^X}
    {\rm exp}\Big( \beta \prod_{\mathsf s \in N_\mathsf i } \sigma_\mathsf s\Big) \,
\end{align}
as expected. Similarly, we can show that $X$ operators are mapped to classical spin operators. Working with the dual representation, i.e. the representation obtained by enforcing the $X$ stabilizer constraints initially, we would have obtained that the norm of the perturbed tensor network maps to $\mc Z_\mathbb A[\beta]$ instead.
Since $\mathbb B_\mathsf i(X) ^2 = {\rm id}$, for every $\mathsf i \in \mc I^X$, a more explicit formula for $\mc Z_\mathbb B[\beta]$ can be obtained as follows:
\begin{align*}
     \mc Z_\mathbb B[\beta]  &=  {\rm tr}(e^{-\beta  \mathbb H_\mathbb B(X)}) 
     = {\rm tr} \Big(\prod_{\mathsf i \in \mc I^X} e^{\beta \mathbb B_\mathsf i (X)} \Big) 
     = {\rm tr}\Big(\prod_{\mathsf i \in \mc I^X} [ \cosh(\beta) \, \text{id}+ \sinh(\beta) \, \mathbb B_\mathsf i] \Big)
     \\
     &= {\rm tr}\Big( \cosh(\beta)^{|\mc I^X|} \prod_{\mathsf i \in \mc I^X}[{\rm id} + \tanh(\beta) \, \mathbb B_\mathsf i] \Big)
     \\
     &= {\rm tr} \Big(\cosh(\beta)^{|\mc I^X|}
     \Big[{\rm id} + \tanh(\beta)\sum_{\mathsf i \in \mc I^X}\mathbb B_\mathsf i + \tanh(\beta)^2 \sum_{\mathsf i_1 < \mathsf i_2 \in \mc I^X} \mathbb B_{\mathsf i_1}\mathbb B_{\mathsf i_2}
     + \ldots + \tanh(\beta)^{|\mc I^X|} \mathbb B_{\mathsf i_1}\cdots \mathbb B_{\mathsf i_{|\mc I^X|}}\Big]\Big)
     \\
     &= {\rm tr}({\rm id})  \, \cosh(\beta)^{|\mc I^X|} \,
     \Big(\sum_{\ell \in {\rm ker} \, \varsigma_X} \tanh(\beta)^{|\ell|}\Big)
     \\
     &= 2^{|\mc S|}\Big(\sum_{\ell \in \ker \varsigma_X} \cosh(\beta)^{|\mc I^X|-|\ell|}\sinh(\beta)^{|\ell|} \Big) \, ,
\end{align*}
where we have used the linearity of  the trace together with ${\rm tr}(\rm id)= 2^{|\mc S|}$. We remark that the term $2^{|\mc S|}\cosh(\beta)^{|\mc I^X|}$ amounts to the partition function of a free model. Hence we deduce from the formula above that there are as many non-trivial contributions to the partition function $\mc Z_\mathbb B[\beta]$ as the number of non-trivial constraints in ${\rm ker} \, \varsigma_X$ satisfied by the operators of the underlying stabilizer Hamiltonian, i.e. products of operators that act trivially on the physical spins. If the number of such constraints does not grow according to the volume of the lattice, we are computing the partition function of a classical model that is essentially free in the thermodynamic limit, and as such it is expected to have a zero temperature phase transition. 
Hence we expect topological tensor network states with global symmetries to be unstable to uniform local perturbations to the tensors.

\end{document}